\documentclass{article}
\usepackage{arxiv}
\usepackage[utf8]{inputenc} 
\usepackage[T1]{fontenc}    
\usepackage{hyperref}       
\usepackage{url}            
\usepackage{booktabs}       
\usepackage{amsfonts}       
\usepackage{nicefrac}       
\usepackage{microtype}      
\usepackage{lipsum}
\usepackage{graphicx}
\graphicspath{ {./images/} }

\title{Detecting cities with high intermediacy in the African urban network}

\author{
 Rafael Prieto Curiel \\
Centre for Advanced Spatial Analysis\\
University College London \\
Gower Street, WC1E 6BT \\
London, UK\\ 
  \texttt{rafael.prieto.13@ucl.ac.uk} \\
   \And
Abel Schumann\\
Sahel and West Africa Club \\
  OECD \\
  2, rue André Pascal - 75775 \\ 
  Paris Cedex 16 - France
  \And
 Inhoi Heo\\
  Sahel and West Africa Club\\
  OECD \\
  2, rue André Pascal - 75775 \\ 
  Paris Cedex 16 - France
  \AND
  Philipp Heinrigs\\
  Sahel and West Africa Club\\
  OECD \\
  2, rue André Pascal - 75775 \\ 
  Paris Cedex 16 - France
}

\begin{document}
\maketitle
\begin{abstract}

Cities play different roles depending on their location within the transport network. Two cities of similar size might have distinct characteristics if one is located on a corridor between two capitals and the other is near a barrier, such as a mountain range. The level of intermediacy is a property of cities that characterises their position in the urban network. We measure the level of intermediacy of African cities by constructing the road infrastructure network obtained from OpenStreetMap. The infrastructure network allows defining city metrics such as degree and centrality. A proxy for the number of journeys that flow through each network edge is approximated using a mathematical model based on the level of attraction or gravity between all pairs of cities. Our model considers the extra time of crossing an international border as a parameter that enables us to proxy the cost of having fragmented regions with costly political barriers. Our results show that small cities have a wide range of intermediacy. We detect a phase transition where cities with less than one million inhabitants have a centrality that depends on the size and degree. For cities above one million inhabitants, centrality tends to be larger and depending primarily on city size rather than degree.

\end{abstract}

\section{Introduction}

{
While there are some commonalities across cities, major differences exist between cities of different sizes and cities in different countries. They are due to their geographical location and role within the national urban system and past policy choices and institutional factors. The spatial distribution of populations and settlements across a country and their interconnectivity and accessibility from urban areas are essential for delivering healthcare, distributing resources, and economic development \cite{50000Criteria}. At a city level, the transport network is often identified as the primary factor that shapes urban patterns \cite{LinardModelAfricaPatterns}.
}

{
Cities are centres of economic activity and transport hubs in all parts of the world. Yet, not every metropolitan area is an equal member of the ``planet of cities'', and many towns and smaller cities are left behind \cite{SecondaryCitiesRoberts}. Outcomes differ especially among secondary cities, some of which are booming, while others are depressed. Yet, secondary cities are and will be home to most of the world's urban population, even if large cities keep growing. Therefore, understanding the social, functional and economic differences of cities, rooted in elements beyond city size, is a critical question for the urban economy. Urban indicators such as high cost of travelling to meet basic needs are one of the main contributors to poverty \cite{LinardPopulationPatterns, 50000Criteria}, whilst distance to hospital shapes the disease mortality rate \cite{Poletti2018}.

Intermediacy is likely to be an essential determinant of the role that cities play within the national economy. Cities with high intermediacy receive considerable through-traffic. They are potentially attractive locations for industries that rely on good accessibility to urban centres and industries that benefit directly from the through traffic. Moreover, the intermediacy of a city has important implications for the role that a city plays within the road network. Cities with high intermediacy are critical within the network. If traffic through them is disrupted, for example, by natural disasters or by armed conflict, the consequences for inter-city travel are high. This is especially important in Africa, where the road network is thin, and few alternative routes exist.
}

{

Although cities with high intermediacy play a critical role in the network, identifying what cities have high levels of intermediacy and what attributes make them more central is a challenge. Among the thousands of cities in Africa, or the hundreds of cities in Ethiopia or Nigeria, some have high intermediacy, but most have not. Two cities with the same population size, even located at a similar distance of a large city, may have vastly different economic outputs \cite{lei2021scaling} and might play a very distinct role in the network, depending on the road infrastructure, the connections to other cities and ultimately, the flow of people, money and of products that pass through each city.

Measuring the flow of products and people that pass through each city and its booming spillover effects in terms of employment creation, for example, with such a level of precision, is a significant challenge, particularly for the small secondary cities. In African cities, data with geographical accuracy is scarce and limited at a continental level. However, precise details are needed to understand the role of cities within the urban network. Detecting the road infrastructure's topology and structure and its proximity to urban agglomerations is a data-intensive task. For example, describing city to city connections through primary roads and motorways from Africa requires considering the sequence of millions of geographical coordinates for describing its road curvature and intricate patterns. Furthermore, there is a complex feedback between the road infrastructure and the distribution of the population across cities. Infrastructure attracts business and more investment, thus becoming a significant pull factor for the population, but also, growing cities are inclined to invest in infrastructure, including new roads, hubs and connections to others. There is an unexplored feedback where cities with high intermediacy attract more infrastructure, increasing their level of intermediacy as a result and attracting more people as well, leaving remote cities behind of that infrastructure. 

}

{
Here, we construct a network of African cities that enables us to measure the level of intermediacy of continental African cities. We leverage open access data from Africapolis \cite{Africapolis} to obtain the location and size of all urban agglomerations with more than 30,000 inhabitants and on data from OpenStreetMap \cite{OpenStreetMap} to construct a connected network of all cities. The fully connected network enables us to compute some country and regional metrics of their road systems. A gravity model is used as a proxy for the number of journeys between every pair of cities. It is based on two assumptions, a distance decay effect and that more people at the origin or the destination increases the number of journeys. Assuming that travellers follow the principle of least effort, each journey between two cities is assigned to the shortest path in the network. The number of trips that travel through each city is then used to compute their level of intermediacy or network centrality in the functional form of a weighted betweenness. Using the time to cross any international border as a model parameter enables us to compute the cost that borders impose on different countries and regions. Detecting sections of the network with high levels of intermediacy enables prioritising planning and designing efficient road interventions, but also highlights vulnerable parts of the road infrastructure, where interruptions due to natural disasters or conflict could have a significant impact. 
}

\section{Identifying cities with high intermediacy} 

City size is one of the most critical aspects that dictate how it functions, its economy, relevance, and role concerning its country and region \cite{pumain2004scaling}. It has been noted, for example, that people from larger cities tend to have a higher income, pay higher rents, create more patents but suffer more crime \cite{GrowthBettencourt} among other socioeconomic and infrastructure aspects of the city that scale with its size \cite{PrincipleSocialScaling}. People from small cities are more likely to move, but in the US, for example, smaller cities are less-preferred destinations for international migrants \cite{ScalingMigrationRPC}. In larger cities, there are many costs and benefits of being surrounded by many people \cite{ScalingInteractions}.

However, city size is only one dimension of complex systems, so other aspects are critical to understanding city life. A city with half a million inhabitants might be a satellite town if it is close to a global city (for example, Ikorodu, with more than 700,000 inhabitants, is a satellite town near Lagos, with 12 million inhabitants) but could also be the largest city within hundreds of kilometres (for instance, Kisangani with more than 400,000 inhabitants in Democratic Republic of the Congo). Beyond city size, other dimensions and relationships with others, such as market potential or level of metropolisation, are more challenging to capture but tend to describe important functional realities of cities \cite{SecodaryCitiesAfrica}.

Many parts of the global and national systems of cities are not being benefited by the new ``Planet of cities'', mainly the secondary cities \cite{SecondaryCitiesRoberts}. Yet, a secondary city might be close to a global city or might be connected to it through some transport system, whilst other cities of similar size might be far not only from big cities but also from the transport network. Two small cities have different industries, accessibility and development opportunities depending on their context with respect to other cities. For example, Shashemene, a city in Ethiopia with 220,000 inhabitants, experiences the flow of thousands of vehicles and tons of products daily, thus becoming an ideal location for industries related to transportation, car repairs, logistics, warehouses, packaging, distribution and other industries which also gain from its strategic location.

\subsection{Cities with different levels of intermediacy}

The intermediacy of cities matters in various context. Most directly, it is a measure of the relevance of a city within a transport network. Thus, it is a useful yardstick for transport planners studying the road network of a country, in particular in countries where systematic traffic flow data is unavailable. High intermediacy indicates particularly important sections of the network. Also, major changes in the intermediacy levels in response to small additions to the network indicate locations where additional road infrastructure can be particularly effective. Intermediacy can also be used to analyse the vulnerability of the road network to interruptions, for example due to natural disasters or armed conflict. Likewise, it can provide a measure of the effects of other disruptions within the network, such as frictions that arise from international borders.
 
Intermediacy as an indicator is especially valuable in combination with other information on the road network, such as data on road quality or traffic flows. For example, differences between actual traffic flows and predicted intermediacy provide information on the economic integration of cities located along a transport corridor. Where actual traffic flows significantly exceed the number of journeys predicted by the intermediacy of a city, the economic and/or social integration of cities located along the transport axis is likely to be exceptionally high.
 
Beyond providing information on the characteristic of the transport network, intermediacy also has important economic implications. Cities with high intermediacy are natural locations for the logistics sector and for other trade related activities. They also benefit from the economic activity that is brought by through traffic, for example for the hospitality sector. At the same time, cities with high intermediacy are likely to suffer from the downsides of high traffic volumes, including air pollution, noise and, depending on whether traffic is routed through the city or around it, congestion.

\subsection{Urban networks}

Considering distinct locations as nodes and the connectivity between them as the edges of a network is one of the most natural ways to analyse spatial data \cite{BarabasiUniversalMobility, barthelemy2014spatial, barbosa2018human}. For example, the map of the metro system of some city contains a minimal set of locations (stations) and connections (lines) expressing all possible journeys within that system \cite{latora2002boston}. Similarly, we can construct an urban network formed by cities and the roads that connects them \cite{wiedermann2016spatial} and detect some properties of the network, such as regular geometric patterns \cite{chan2011urban}. Considering an urban network enables us to import many concepts, metrics and algorithms from network theory, such as \emph{connectivity}, \emph{node degree}, \emph{shortest path} or \emph{centrality} \cite{BarabasiStructureNetworks, SmallWorldStrogatz, RevealingCentrality} for a set of cities and roads.

Spatial networks, as opposed to other types of networks, have interesting properties since objects are embedded in space, so there is a natural distance associated with edges \cite{gastner2006spatial, barthelemy2011spatial}. At a city level, for example, a spatial network made of streets, \cite{crucitti2006centrality, buhl2006topological}, roads \cite{lammer2006scaling} or metro stations \cite{latora2002boston} has been used to study the structure of a metropolitan area and detect some spatial patterns, such as centrality, efficiency or hierarchy \cite{barthelemy2019statistical}. At a bigger spatial scale, the network of the Korean highway system \cite{jung2008gravity}, or European cities \cite{dupuy1996cities}, US metropolitan areas \cite{garrison1960connectivity}, or the network of Chinese intra-urban mobility has been constructed \cite{lei2021scaling}, where the nodes of the network are the cities. The connections are usually based on some spatial proximity between the nodes or infrastructure, such as roads or highways \cite{kansky1963structure}, or even interurban commuting traffic, where complex features and nontrivial relations with the underlying topology properties of the network can be observed \cite{de2007structure}.

As with every model, certain simplifications and assumptions are made, favouring other aspects. In the case of urban networks, cities are usually simplified and represented by the nodes of a graph, and the roads are the edges. Thus, many assumptions and simplifications are made, for example, to decide which roads and which cities to keep, define when a road passes through a city, or detect when two roads are connected. In turn, a network is obtained, that gives a structured manner to handle large and complex data using only two sets: nodes and edges. It then becomes possible to measure a city's degree, its centrality or to detect the shortest path between cities.

We aim to investigate the level of intermediacy of a city based on the number of journeys that pass through it. Based on the urban network, we first model the number of journeys between each pair of cities and then assign those journeys in the network. Assigning some trips given some origin and some destination with an unknown route is a frequent problem in traffic models. In general, some origin-destination data is available, and the challenge is to assign journeys to routes \cite{wardrop1952road, daganzo1977stochastic}. Furthermore, once assignments are made, it is possible to measure the level of centrality of different nodes, which represent not the number of connections but a more complex relationship between a node and the rest of the network \cite{crucitti2006centrality, barthelemy2018morphogenesis}.

\subsection{A gravity model to estimate journeys between cities}

The gravity model is one of the most prominent models for social mobility and for considering interactions between distant spatial units. Whether we apply it for mobility, migration, trade, or others, the gravity model for human constructs should be thought of as a general and very flexible model, which captures the impact of size at the origin and destination and their distance \cite{jung2008gravity}.

The gravity model usually depends on the size of the origin, frequently captured by its population. The flow of people, the demand for some product or the number of migrants is expressed as some increasing function of the size of origin, assuming that with more people, then a higher demand or a higher flow is expected. The gravity model captures the \textit{attractivity} of the destination and, in general, utilises population size as a proxy. However, the model can be applied to other dimensions, such as GDP or the number of hotel rooms, to capture the size of some touristic destinations, for example. In terms of the distance, the actual physical distance between origin and destination is frequently used, but the model allows other forms of pairwise relationships that might capture proximity, such as travel time or other aspects, including frictions or barriers \cite{Gravity, GravityModel, ZIPMigration, NorthAmericaMigration}. The gravity model, for example, has been used to model trade between countries, considering the population (or income) size and the physical or cultural distances between two distinct locations \cite{GravityModel, Gravity}. 

The gravity model has many expressions, depending on the type of data available and on its application. One of such expressions is given by
\begin{equation} \label{Gravs}
F_{o,d} = \kappa \frac{P_o^\mu P_d^\nu}{D_{o,d}^\gamma},
\end{equation}
where $P_o$ and $P_d$ are the population size of origin and destination. The numbers $\mu$ and $\nu$ are model parameters that capture non-linear impacts of city size. The denominator has $D_{o,d}$, which is the network distance between the origin and destination, and $\gamma$ captures the impact of distance. Finally, $\kappa$ is a constant to adjust the observed units. The expression, with $\mu = \nu = 1$, with $\gamma = 2$ and some (fixed) value of $\kappa$ resembles Newton's law of universal gravitation, thus the name ``gravity'' for social models. However, non-linear effects that scale with city size were detected by looking at people who move and the destination picked. A sublinear relationship with city size was observed on the probability of moving in the US \cite{ScalingMigrationRPC}. Also, the value of $\gamma = 2$ has been observed in different settings \cite{barthelemy2011spatial, schlapfer2021universal, MobileCommunication}, so it is frequently used, but there is no reason why other values should not be considered.

Parameter estimation of the gravity model is usually feasible with enough observed data \cite{GravityR}. That is, the parameters $\mu$, $\nu$, $\gamma$, and $\kappa$ might be estimated in equation \ref{Gravs} if enough pairwise flows (left-hand side of the equation) are known, given the corresponding sizes and distances. A frequent technique (although problematic with too many zero flows) is to take the logarithms on both sides of equation \ref{Gravs} and estimate the parameters and intervals through a regression. However, other estimation techniques, based, for example, on a Poisson regression or other statistical methods, are also possible \cite{ZIPMigration, GravityPanelData}.

Here, however, we have the opposite situation. The size and location of each metropolitan area are available, and the commuting times between each pair of cities is estimated. Yet, limited information is known concerning the actual number of journeys (or trade, or interactions or financial transactions) for different pairs of cities. Here, the gravity model is used in its functional form, that is, we consider some values of the four parameters, $\mu$, $\nu$, $\gamma$ and $\kappa$, we input the city size of origin and destination, and the distance between them on the right-hand side of equation \ref{Gravs}, and we obtain some modelled flow as a result, on the left-hand side of that equation. 

\subsection{Measuring city centrality}

There are many ways to measure a node's importance in a network, for example, by its degree \cite{freeman1978centrality}, or a combination of its degree and the degree of neighbours \cite{bonacich1972factoring, buechel2013dynamics}, or a weighted combination of node and edge attributes \cite{singh2020node}. Node betweenness is a commonly used metric that gives the number of times that a node occurs in the shortest paths between all possible pairs of nodes in a network \cite{freeman1978centrality}. However, we are interested in the shortest paths between all possible pairs of nodes, but for each pair, we will estimate the number of journeys between them using a gravity model.

It is possible to consider node weights for computing the centrality of the node based on the same idea of node betweenness \cite{singh2020node}, whereby we count the number of times that a node occurs in the shortest paths between all possible pairs, but instead of a simple count, we add the gravity between origin and destination. Here, centrality is constructed for each node as the sum of the number of journeys that pass through a city, or simply the ``flow'' through a city, referring to people, products, transactions, cash or any other good.

\section{Methods}

\subsection{Data}

Primary roads, highways and trunks were obtained from OpenStreetMap. The latest Africa OpenStreetMap data was downloaded on March 17, 2021 from http://download.geofabrik.de/. The data gives the $x,y$ coordinates as a sequence of vertices of each road segment. Other road types, including paths, secondary and tertiary roads, are not considered since most are part of urban polygons and are not frequently used for city to city mobility. The raw data contains 5.4 million coordinates that form 221,378 road segments, with 415,231km of roads in the continent. 

For straight roads, fewer coordinates are needed, but curves are composed with a sequence of multiple vertices. We assume that a straight road connects them for every two consecutive vertices, and we keep the physical distance between the two vertices as the length of that segment, using \cite{citeGeosphere, citeR}. For each road, the sum of all the segments gives us the road length. If a road is curvy, its length is approximated by the sum of all the segments, thus providing a good estimate of the road length.

\subsection{Constructing the urban network}

Since we are interested in constructing a cities network, we begin the process with the coordinates of all urban agglomerations with more than 100,000 inhabitants. Data from Africapolis \cite{Africapolis} gives the location of 705 cities. Each city is then assigned to its closest road coordinates. For most cities, the nearest road segment is within a few metres, and it goes up to 12km. Some of the transport nodes are close to a small city (with less than 100,000 inhabitants). If the distance between a city and a transport node is less than 10km, we assume that the road passes through that city, so we label the closest transport node as a city node. Here, 1,457 cities with less than 100,000 inhabitants are added to the network nodes. The network has 2,162 cities, including all cities with more than 100,000 inhabitants and smaller towns close to the road infrastructure. 

Each coordinate from the road infrastructure is labelled as a \emph{city}, if it corresponds to one, as \emph{terminal} node if it is starting or ending point of the road, as \emph{crossing} if there are more than two options on that node (so it includes bifurcations and road crossings) and simply as \emph{road} nodes if it is neither.

A first graph is constructed using all the labelled vertices as nodes, and the corresponding road segments are the network's edges, using \cite{citeR, citeIgraph}. For each edge, we keep the type of road and its length as edge attributes. This first network is not connected (as there are many disconnected patches in the data), and it has 36,000 nodes, 220,000 edges with 3,500 disconnected components. The following two steps connect and simplify the network.

\subsection{Connecting and simplifying the network}

The first step is to create a fully connected network. Data from OpenStreetMap often have highways with a gap in between (a gap that could be a few metres). Hence, we first explore all terminal nodes. If a terminal node is less than 24km away from any other road segment, we assume it is possible to travel between them, so we add that edge. Since that edge did not exist in the OpenStreetMap data, we labelled it as ``added''. Two terminal nodes might become road nodes if they get connected, but also, a road node might become a crossing if a new travel option is added. The network has no terminal nodes at a short distance to others by the end of this procedure.

Secondly, we check whether there are any disconnected components in the graph. We identify the physical distance between all the nodes in one of the components against every node in the remaining part of the graph. We add the corresponding edge with the shortest distance and label it as ``added'' as well. For all the added edges, we cannot guarantee that travelling through them is feasible or efficient (fast), but the result is a fully connected network. We can then estimate the network's travel distance between all pairs of nodes and whether the edges are primary roads, highways or if our procedure added them.

Finally, for every pair of nodes with a physical distance smaller than 24km, we measure their road distance (the sum of the lengths of all the segments of the shortest path between them) and compare it against its physical distance. If the ratio is greater than 10, we add the edge between them and label it as ``added''.

We simplify the network by removing (dissolving) road nodes with a degree of two that connect the same type of road at both ends and replace it with a ``longer'' edge. We sum the lengths of the corresponding lengths to form the new edge so that instead of having a long sequence of road nodes, we finish with a network with one less node and edge. Also, we remove all loops, that is, edges that begin and end in the same node. The procedure gives in a network that has cities, crossings or terminal vertices as the nodes and has the travel length and the type of road for each edge that connects the nodes. The result is a fully connected network with 7,361 vertices, where most of them (71\%) are road nodes and 9,159 edges (Figure \ref{TypesOfNodesEdges}).

{
\begin{figure} \centering
\includegraphics[width=0.7\textwidth]{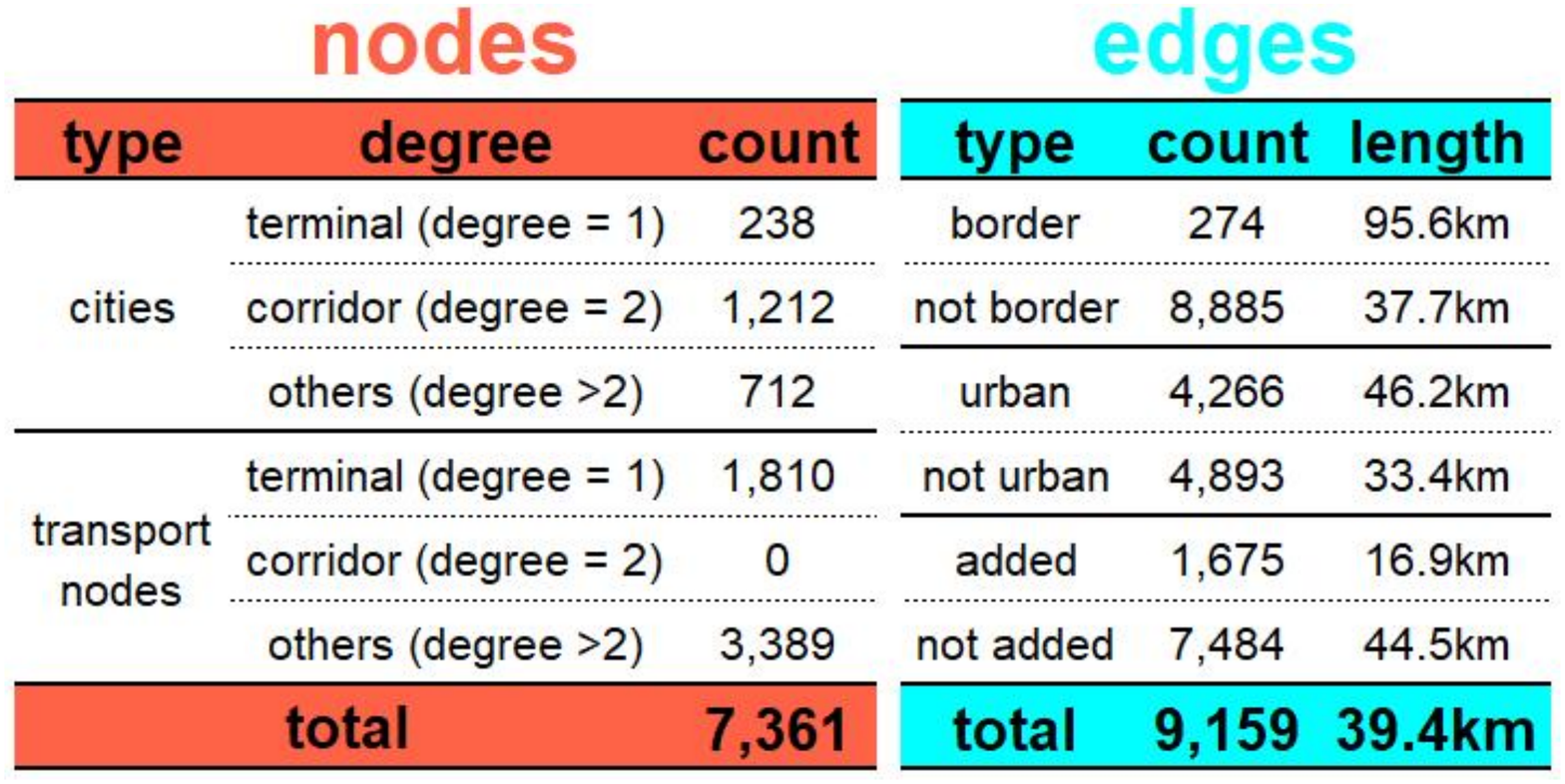}
\caption{The simplified network has 7,361 nodes and 9,159 edges.} \label{TypesOfNodesEdges}
\end{figure}
} 

Data from OpenStreetMap in Africa contains more than 220,000 primary roads, motorways and highways. Each road is made up of a sequence of coordinate points that defines a polygon with its curvature. On average, each road comprises 24.2 points (with a maximum of 1,999 points for a single road), meaning that the road data has 5.4 million points in its original form. Our method reduced the 5.4 million points (that in some cases are disconnected polygonal lines or patches) into a connected network with only 9,159 edges and 7,361 nodes. The labelled network contains information about the length of each road segment (except for the added ones, for which its length has to be estimated). Instead of working directly with the road coordinate points, we have a network where computing distances between nodes and local or global metrics such as centrality or degree is now possible.

For spatial networks, it is possible to measure the Euclidean distance between nodes (``as the crow flies'' distance), but also the total distance of the shortest path between any two cities \cite{barthelemy2011spatial}. Here we have additional information from the length of each segment, the type of road and whether it crosses a border.

\subsection{Travel times}

We transform travel lengths into travel times. The motivation is to compare different types of road (so that travelling 10km on a highway is faster than travelling through added edges and roads) and because travel time enables us to add an extra cost for crossing an international border. Any edge with its extremes on different countries is a border crossing, so we add additional $\tau$ minutes for crossing it. With $\tau=0$, there is no time cost on crossing an international border, but higher values mean that the edge is less efficient.

We also add extra travel times for crossing any urban agglomeration. Suppose any of the extremes of an edge is in a city with a population $P_i$. In that case, we assume that the travel time of that edge has an extra cost (from traffic and from crossing an urban environment) that is proportional to $\theta \sqrt{P_i}$. The motivation is to capture the city's surface with the population, so the radius inside that city approximates the extra travel time. This way, with $\theta = 1/30$, the extra travel time from crossing a city like Cairo, with 22 million inhabitants, is additional 160 minutes, but the extra travel time from crossing a city with 100,000 inhabitants is 10 minutes, so nearly 16 times more cost from crossing Cairo than a small city.

The result is a fully connected network, where for each edge, we capture the type of road, its length, whether it is an international border or if it goes through an urban environment. We compose it all into some estimated travel time.

\subsection{Shortest travel time path}

For every pair of cities, we estimate the shortest path that connects them in terms of the travel time and compute the travel time between them. It is assumed, based on the principle of least effort, that any person travelling between these two cities will choose the shortest path between them \cite{Zipf1949LeastEffort}.

The parameter $\tau$ that indicates the extra time for crossing an international border modifies the costs of some of the edges, and so it also alters the shortest paths. Still, for every value of $\tau > 0$, it is possible to compute the shortest path between every pair of nodes.

\subsection{Calibrating flows between any pair of cities}

For some value of $\tau > 0$, the travel time between two cities is estimated and used as input in equation \ref{Gravs} as the distance between them. Although asymmetric flows and some scaling behaviour could be observed, we assume that $\mu = \nu = 1$ so that the flows scale linearly with city size of origin and destination. Further, different values of $\kappa$ capture total flow within some time unit, so it is possible to adjust units and intervals and assume that $\kappa=1$. We then have a modified version of equation \ref{Gravs}, given by
\begin{equation} \label{GravsSimple}
F_{o,d} = \frac{P_o P_d}{ND_{o,d}^\gamma},
\end{equation}
where $ND_{o,d}$ is the travel time in the network between the two cities. Then $\gamma$ is our model parameter that indicates the impact of the network distance. With $\gamma=0$, for example, the distance has no impact, and the estimated flow between any two cities only depends on city sizes. However, with larger values of $\gamma$, the flow between distant cities becomes negligible, and we only observe local flows (so we get interactions mostly between nearby cities).

For two cities in West Africa, Tamale and Ouagadougou, the cumulative percentage of products that come within a certain distance was measured \cite{FoodshedsTwoAfricanCities}. We use the observed cumulative trade at a certain distance for these two cities to approximate our parameter $\gamma$ value. We minimise the square error observed at distinct distance thresholds. Our results show that with a value of $\gamma = 2.8$, the cumulative product intake from Tamale and Ouagadougou is as close as possible to the observed ones. The modelled cumulative curves suggest that higher parameter $\gamma$ values force the trade to be more local from nearby cities. More details about the estimation of $\gamma = 2.8$ in the Supplementary Materials.

\subsection{Estimating the number of journeys through each city}

We estimate the number of journeys that travel between each pair of cities using equation \ref{GravsSimple} by considering the size and network distance on the right-hand side, and then we assign that flow through the shortest path in the network. We add the number of journeys that pass through each node in the network when all pairs of cities are considered. 

We obtain an estimate of the number of journeys $j_k$ that pass through node $k$ during a certain period. Small values of $j_k$ suggest that only a few trips pass through that city, so most of its flow is originated only from its population. Large values indicate that many trips pass through that city, even when they did not arise there. Thus, a city with large values of $j_k$ has a strategic location in the network that connects distinct clusters or a large city size to compensate.

\section{Results}

We first analyse the fully connected and simplified network (Figure \ref{NetworkNoNames}). In total, we consider 2,162 cities with a total population of 460 million inhabitants (roughly half of the continent's population). The remaining population lives in smaller cities and rural areas.

{
\begin{figure} \centering
\includegraphics[width=0.9\textwidth]{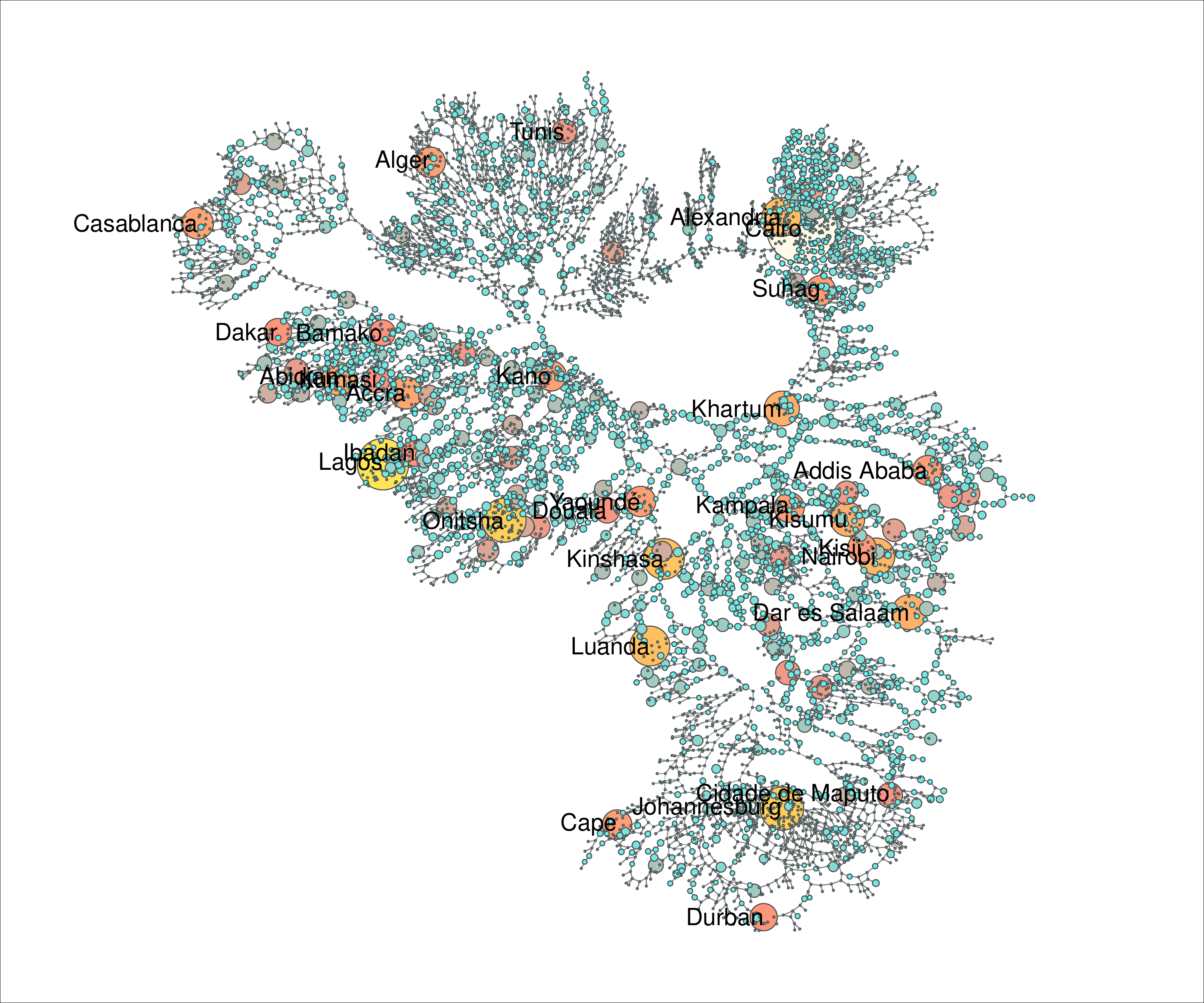}
\caption{Network of African cities. The size of the disc is proportional to city size, and the width of the edge is inversely proportional to the length. The location of each node is slightly transformed from its physical coordinates for a visual representation of the network so that the upper part of the graph corresponds to North Africa and similarly for other regions.} \label{NetworkNoNames}
\end{figure}
} 

\subsection{Degree per city}

The degree of a city suggests its role in the network. For example, 4\% of the cities have a degree of one, so they are terminal nodes. Most of the terminal cities (60\%) have a population smaller than 100,000 inhabitants. Cities with a degree of two can be thought of as corridors or an agglomeration that grew around some road. Most of the cities are corridor cities, with a degree of two (Figure \ref{DistDegree}). Corridor cities also tend to be small in population (83\% of the corridor cities have less than 100,000 inhabitants). Cities with a degree of five or more are, in general, large metropolitan areas with an average population of 1.2 million inhabitants.

{
\begin{figure} \centering
\includegraphics[width=0.6\textwidth]{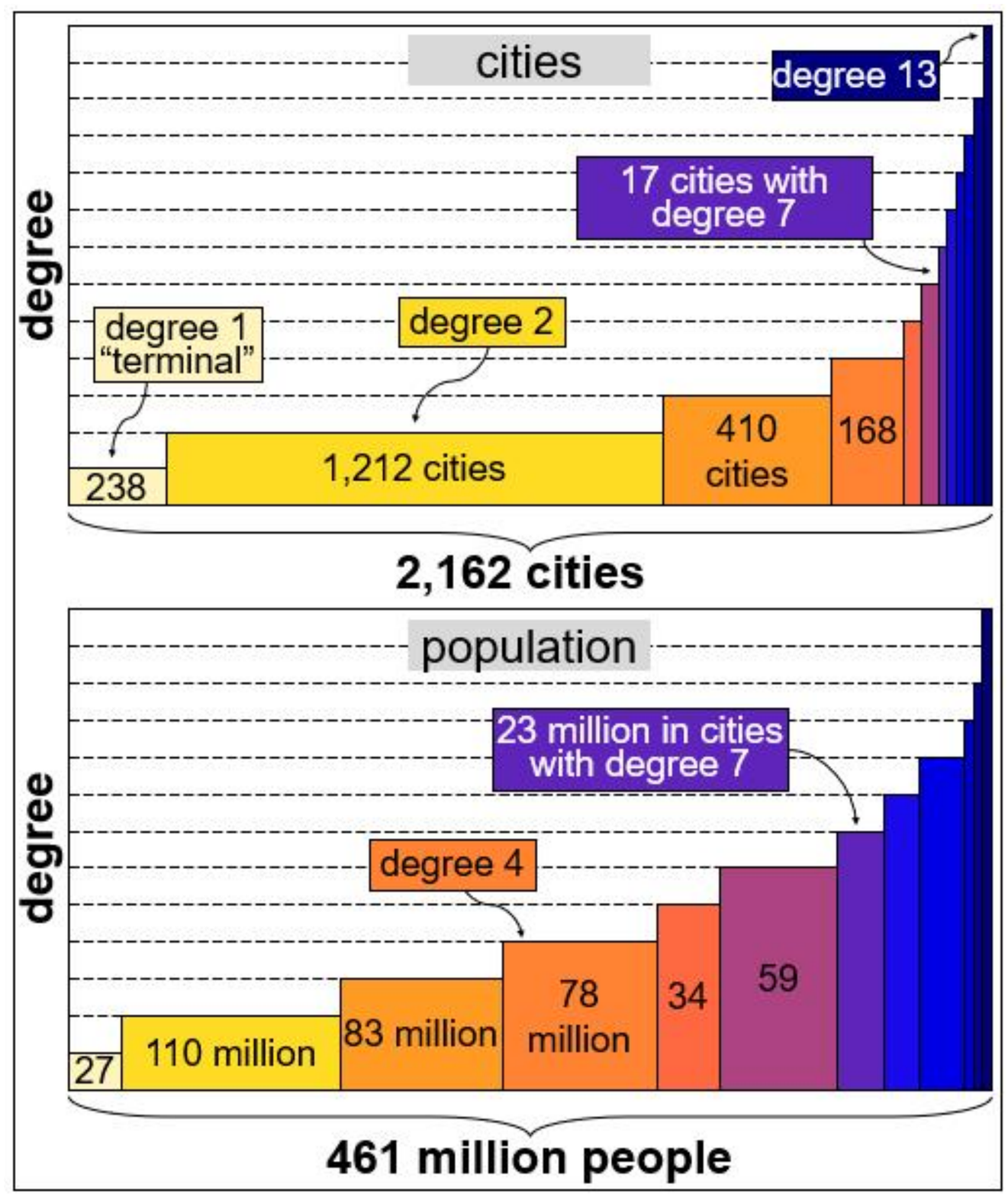}
\caption{Degree of each city (top) and the corresponding population. There are 1,212 cities with degree of two (so, a corridor, with two entries). Corridor cities tend to be small, so in 1212 cities (56\% of the cities) there is only 109.9 million individuals (23.8\% of the population considered).} \label{DistDegree} 
\end{figure}
}

We find that 80\% of Ethiopian cities are corridor cities, 79\% of cities in Somalia and 75\% of cities in Mauritania, suggesting a more simple network than South Africa, for example, where 49\% of the cities are corridor cities. Only 22\% of cities in South Africa with more than 100,000 inhabitants are corridor cities, and one in four cities in Egypt is a corridor city. Also, the average degree of cities suggests the complexity of its infrastructure network. For example, from cities with more than half a million inhabitants, the average node degree gives 8.6 in Algeria and 6.3 in Tunisia, but 2.5 in Benin and 2.0 in Somalia and Tchad.

Another relevant aspect at a country level is the abundance of transport nodes. Describing the transport network in Egypt requires 625 transport nodes, in South Africa 719 nodes, and 752 in Algeria. But in countries of similar size and population, the network is much less complex. In Ethiopia, only 80 transport nodes are needed, 79 in Kenya, 88 in Ghana, and 41 in Sudan. We find 11.2 transport nodes per million urban population in Africa, with only 1.0 transport nodes per million in Somalia, suggesting a more simple network.

From the 935 small cities in the network (urban agglomerations with less than 100,000 inhabitants), 63\% of them have a degree of two and 25\% have a degree larger than 2, corresponding to agglomerations located in some road crossing or bifurcation. From these small cities and only due to their connections in the network, those with a degree of three or above are 20\% more central than cities with a degree of 1.

\subsection{Scaling of node degree per city}

The simplified network shows that the node degree scales sublinearly with city size (Figure \ref{ScalingDegree}). That means that the degree $D_k$ of city $k$ increases with city size, according to $D_k = \alpha P_k^\beta$, with $\hat{\alpha} = 0.2878 \pm 0.0278$ and $\hat{\beta} = 0.1816 \pm 0.0085$, meaning that larger cities tend to have a higher node degree, but the degree of the city increases very slowly with more people. For example, if we compare a city with one million inhabitants against a city with ten million, keeping everything else equal (including geography and legacy), then the large city has a degree $10^0.1816 \approx 1.5$ times larger.

{
\begin{figure} \centering
\includegraphics[width=0.7\textwidth]{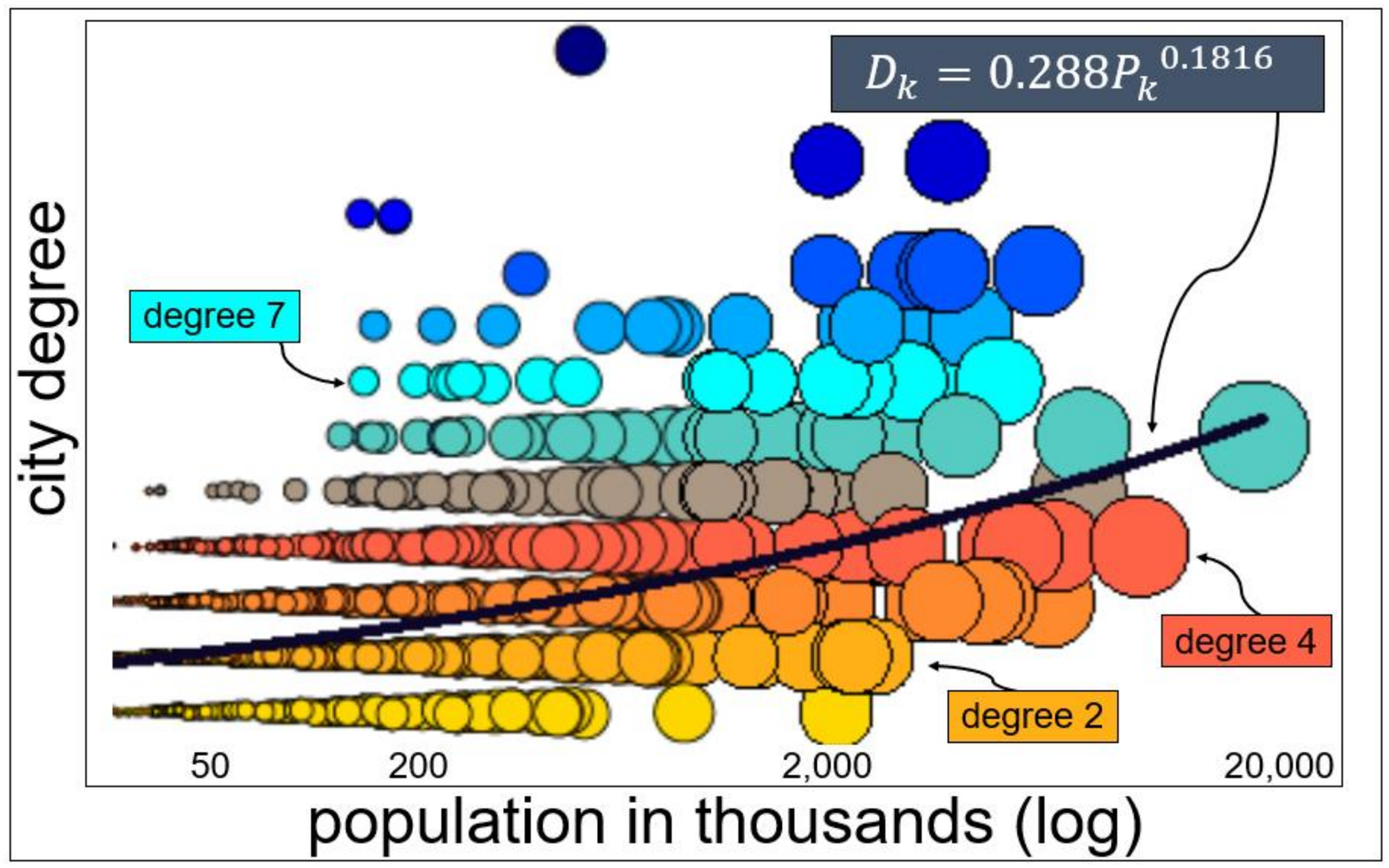}
\caption{Degree of cities (vertical axis) according to the population (horizontal, log scale). The degree of a city scales very sublinearly, so a city has to increase its size 45.4 times to double its degree.} \label{ScalingDegree} 
\end{figure}
}

\subsection{Travel distance}

The network distance between any two cities in Africa is, in general, 1.38 times the physical distance between them \ref{PhysicalVsNetworkDistance}. There are two reasons for this ``extra'' network distance. Roads are generally not straight, but also, a scarce infrastructure means that journeys are detoured through the available edge segments. For these reasons, the network to physical distance ratio varies across countries and regions.

{
\begin{figure} \centering
\includegraphics[width=0.8\textwidth]{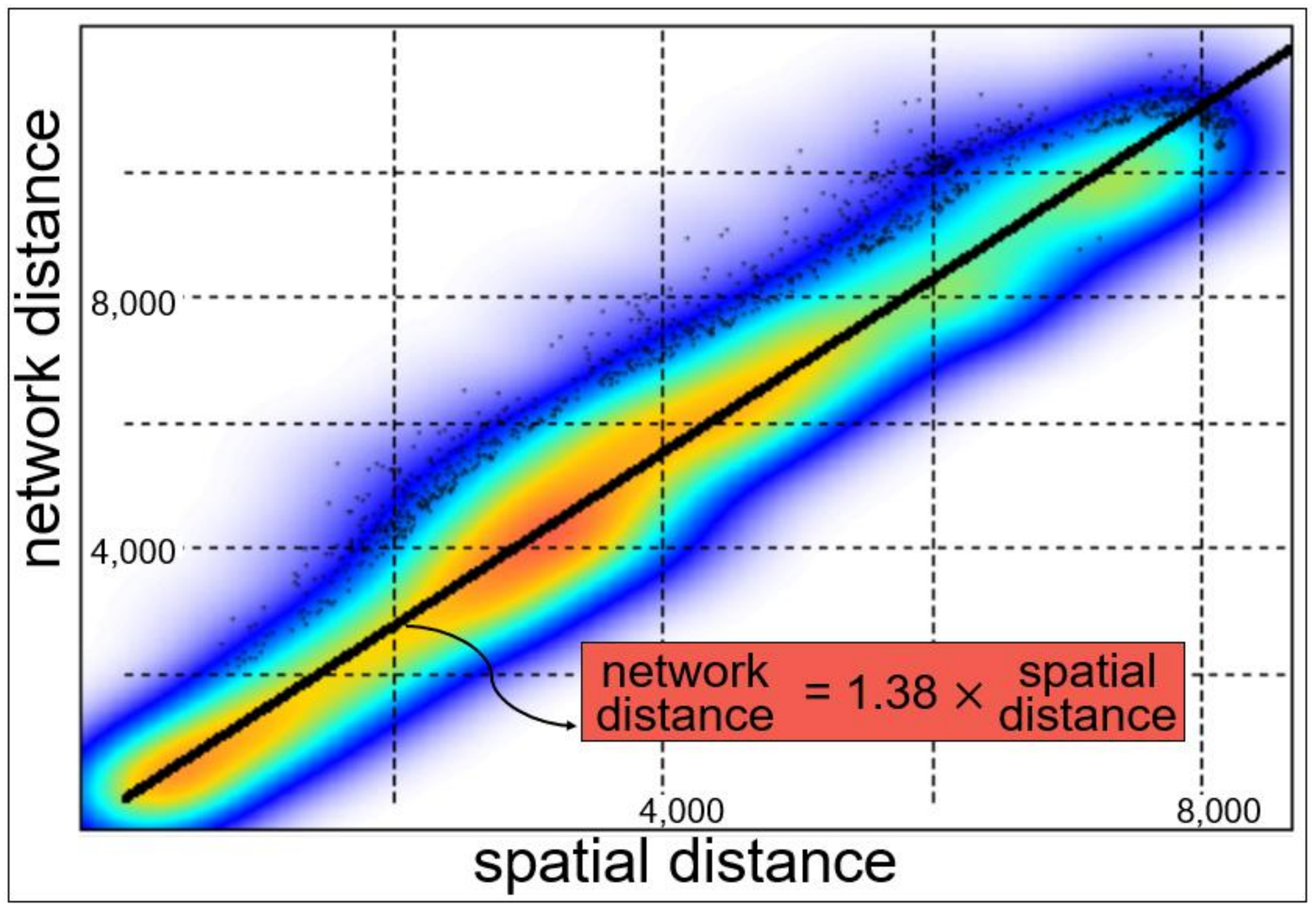}
\caption{Spatial distance (horizontal axis in kilometres) between each pair of cities, and network distance between each pair of cities (vertical axis). On average the network distance is 38\% larger than the spatial distance.} \label{PhysicalVsNetworkDistance}
\end{figure}
}

We measure the \emph{curvature ratio} of a road by comparing its length (our proxy for the road length in kilometres) to the spatial (geodesic) distance between the starting and ending point. If the ratio is close to one, the road is approximately a ``straight'' line, but a larger ratio means higher curvatures on its segments. 

Yet, even if roads are mostly straight, the network distance between two cities might be large (or at least more significant than their spatial geodesic distance) if there are not many direct roads between pairs of cities. We construct two country-level metrics to capture its roads' curviness and the abundance (or lack) of road infrastructure. Firstly, the country road curviness is defined as the average road curviness for all roads that start and end in that country. Secondly, to measure infrastructure from a country, we consider all pairs of cities from that country and measure the average ratio between the network and the spatial distance (Figure \ref{SimulatedNetwork}).

{
\begin{figure} \centering
\includegraphics[width=0.8\textwidth]{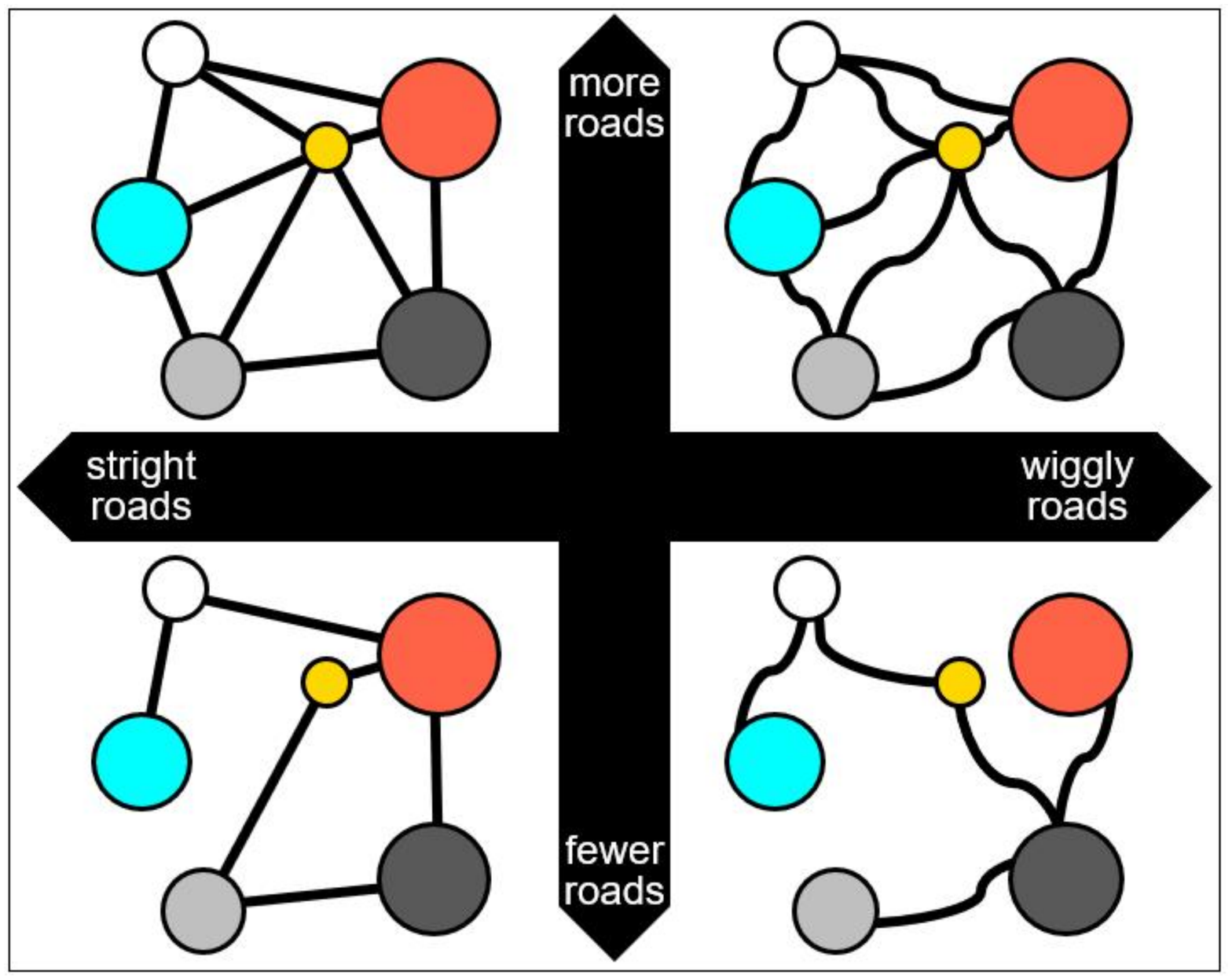}
\caption{Two attributes from the road infrastructure at a country level are observed. Firstly, if the roads are straight or curvy (horizontal) and secondly, if there are many roads to connect cities. } \label{SimulatedNetwork}
\end{figure}
}

Our results show that in Egypt, for example, there is some road curviness, but still, the average distance between each pair of cities is close to one, meaning that in the country, the spatial distance is a good approximation of the network distance (Figure \ref{ScatterRoadMetrics}). In Mali, roads are curvier, and so road distances are also higher. In Rwanda, roads have the highest average curvature, and the network distance almost doubles the average spatial distance.

{
\begin{figure} \centering
\includegraphics[width=0.8\textwidth]{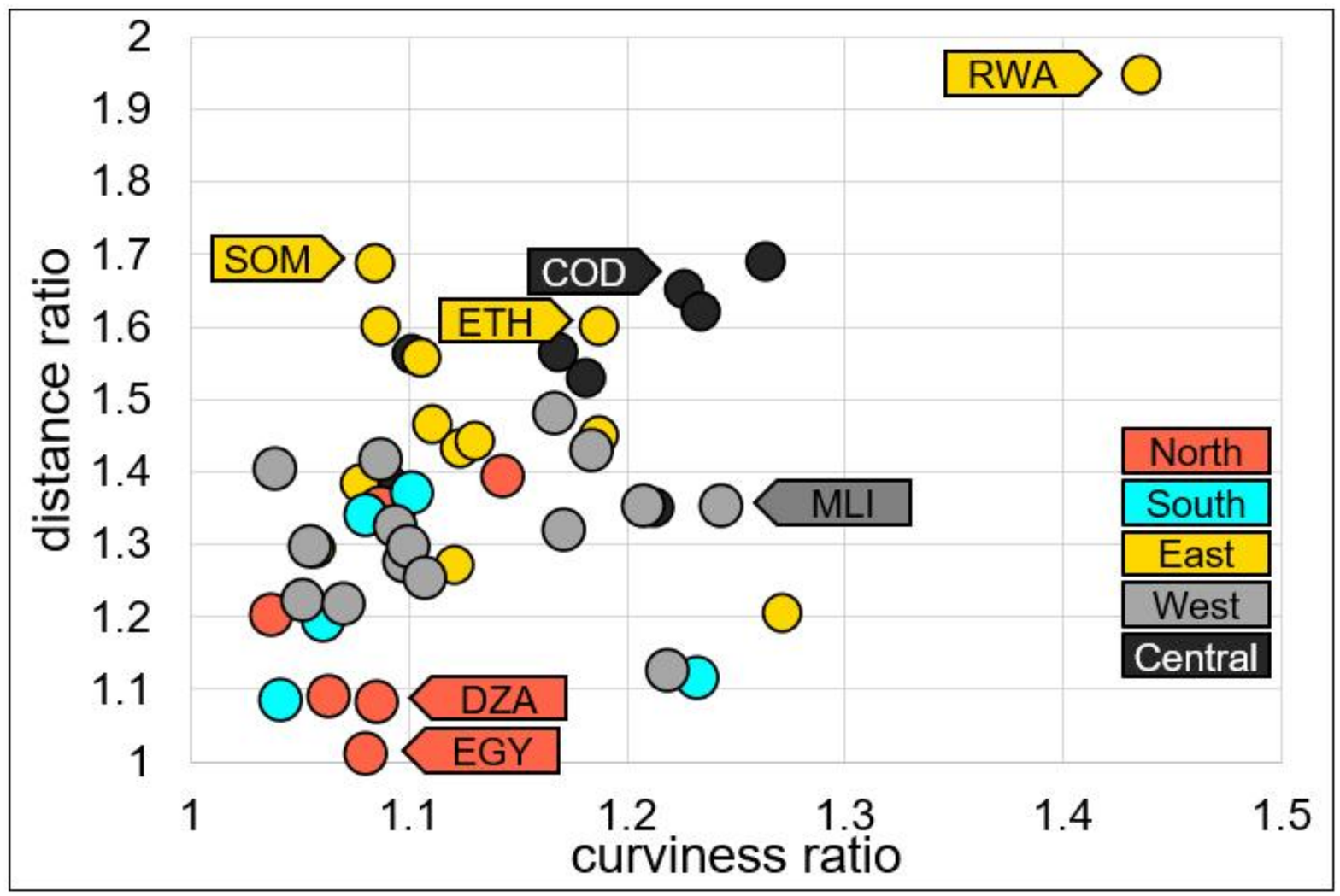}
\caption{Average road curviness in each country (horizontal axis) and average ratio between network distance and spatial distance (vertical axis).} \label{ScatterRoadMetrics}
\end{figure}
}

\subsection{Cities with high intermediacy}

The centrality increases with city size (Figure \ref{CentralityScatter}). Still, for cities with 200,000 inhabitants, for example, there is a large variation in the level of centrality. Some cities have the centrality levels of large cities, such as Lagos or Johannesburg, even if their population is 100 times smaller. Cities with a high centrality but a small population tend to be national and state capitals, enjoy a strategic location in the network and form a crucial element of the national urban system, as they function as transport hubs.

{
\begin{figure} \centering
\includegraphics[width=0.8\textwidth]{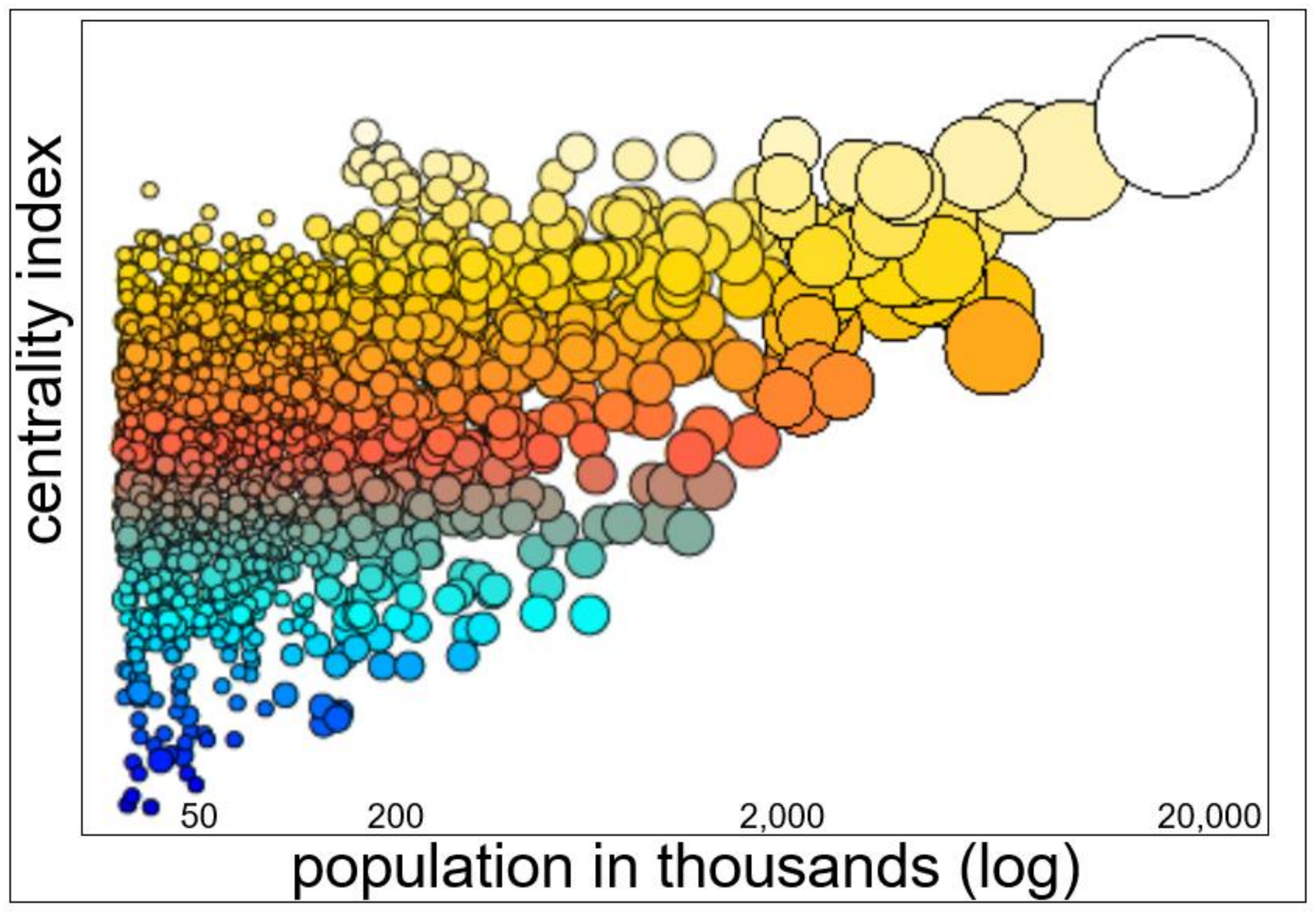}
\caption{Centrality (vertical axis) and population (horizontal axis) of the 2,162 cities considered in the African network of cities. } \label{CentralityScatter}
\end{figure}
} 

Cities with high centrality have a strategic location in the network, with some proximity to large cities and likely with higher node degree (Figure \ref{CentralityMap}).

{
\begin{figure} \centering
\includegraphics[width=0.8\textwidth]{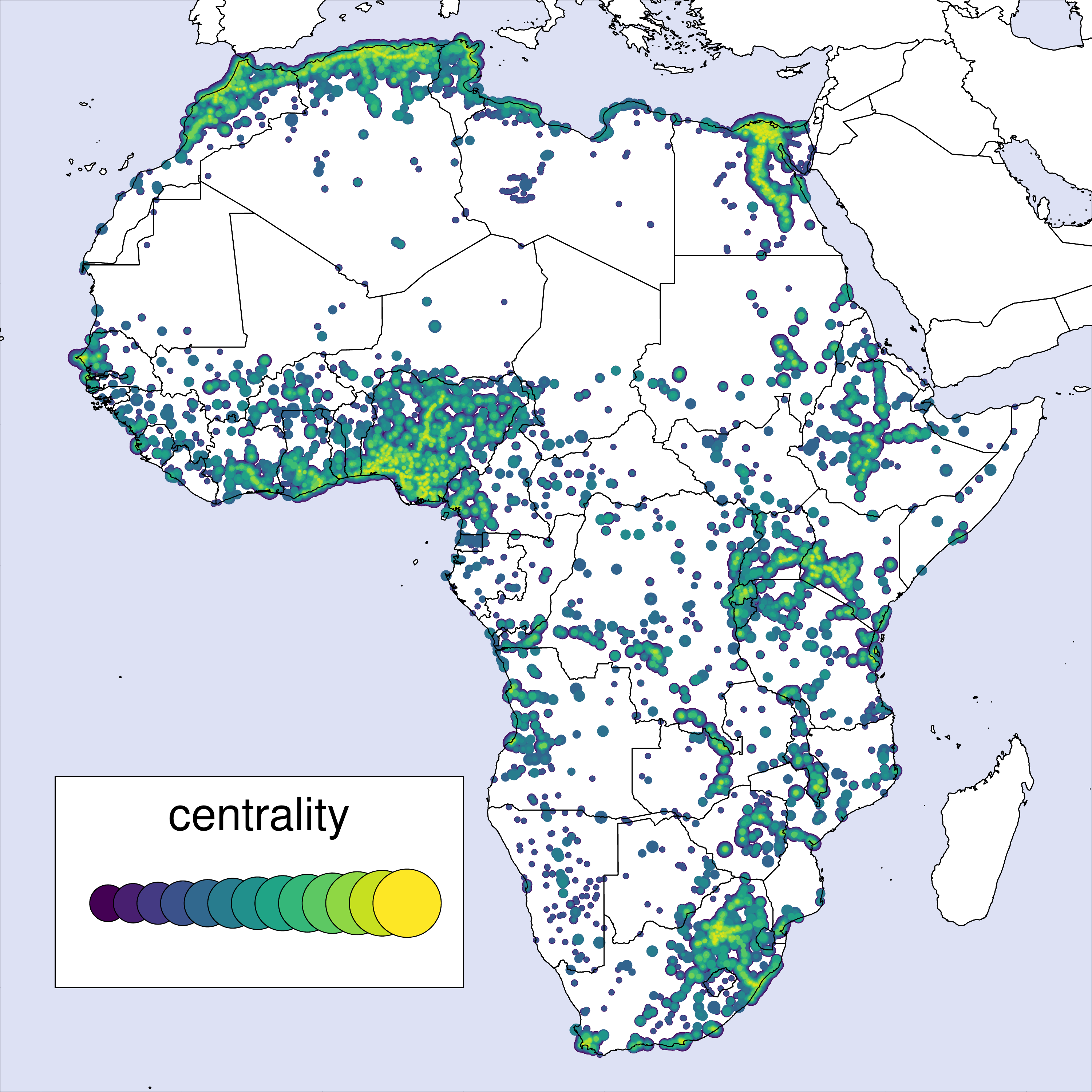}
\caption{Level of centrality of all cities. More central cities are bright, less central cities are dark red.} \label{CentralityMap} 
\end{figure}
} 

The centrality of transport nodes is also a reflection of the complexity of the network. From the 5,199 transport nodes, all with a degree of one have a centrality of zero, since no journey minimise their distance from a city to another passing through a terminal node. Also, there are no transport nodes with degree 2 since they were dissolved in the network. But for transport nodes, any extra degree increases the centrality on average by 38\%, so more central nodes are also those with a high node degree.

\subsection{The cost of international borders}

We compute the total trade for each country for different border costs $\tau>0$ and compare the trade against the case with $\tau = 0$, so the case with no additional time for crossing a border. Results show that depending on the country, the impact of borders can be quite significant (Figure \ref{BorderCostsWAfrica}). For example, in West Africa, the total trade of Benin and Togo goes from 100 units to 70 units if the border costs are two hours or more. The costs for other regions in the Supplementary materials.

{
\begin{figure} \centering
\includegraphics[width=0.8\textwidth]{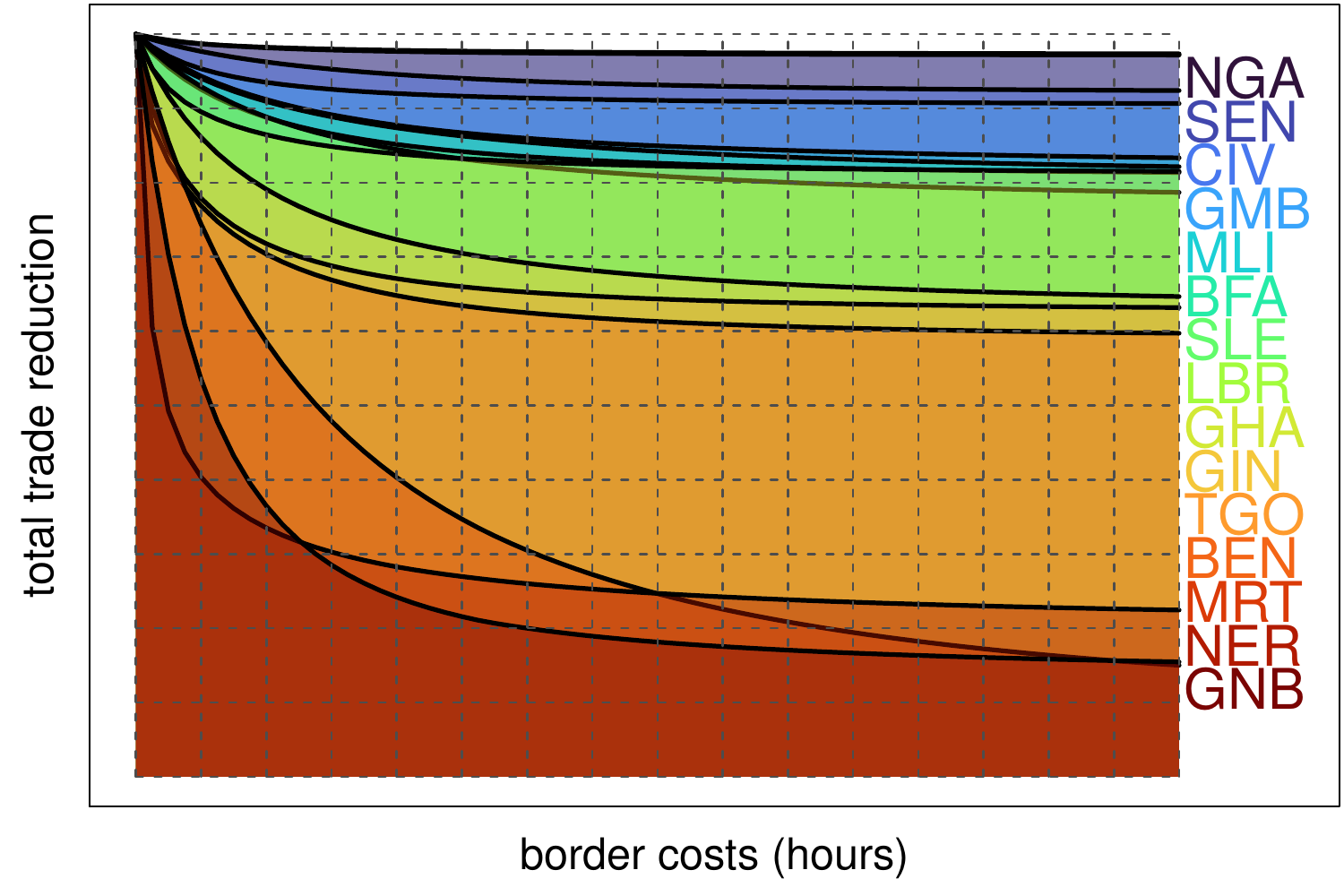}
\caption{Costs of international borders $\tau$, in hours (horizontal axis) against the ratio between the the total trade and the trade with no border costs (vertical axis) for West Africa.} \label{BorderCostsWAfrica}
\end{figure}
}

\subsection{What is a city with high centrality}

Shashemene, in Ethiopia, is a city with 220,000 inhabitants (ranked 11 in Ethiopia and 76 in East Africa) is a secondary city with a high centrality (ranked 3 in Ethiopia and 16 in East Africa). Shashemene has a degree of five, second largest degree in the country and fourth in East Africa (only the capital Addis Ababa has a degree of nine and Harare and Kampala have a higher degree than Shashemene in East Africa) and has a strategic location in a North-South and East-West corridor. On the opposite side of the spectrum, Yirga Chefe has a higher population (with 270,000 inhabitants is ranked 9 in Ethiopia) but, being far from a big city or a large corridor, and with a degree of two (so, a corridor city), it has only 21\% of the centrality of Shashemene. The degree and location of a city in the network are critical components beyond city size that enhance or hinder the city's centrality.

We construct a multivariate regression with the number of journeys through each city as dependent variables and city size, degree, and region as independent variables. Results show two regimes. For small cities, we observe that the degree and city size are critical covariates that determine city centrality (Figure \ref{WhatIsHighCentral}). There is a phase transition for cities at around one million inhabitants, where city size becomes the most prominent variable, although still city degree plays some minor role. The number of connections from the city for cities with two or even five million inhabitants is less significant, and centrality is mainly dominated by city size. Large cities (above five million inhabitants) tend to have a high city degree (Figure \ref{ScalingDegree}). However, for large cities, its size is a determinant factor in its centrality levels. 

{
\begin{figure} \centering
\includegraphics[width=0.8\textwidth]{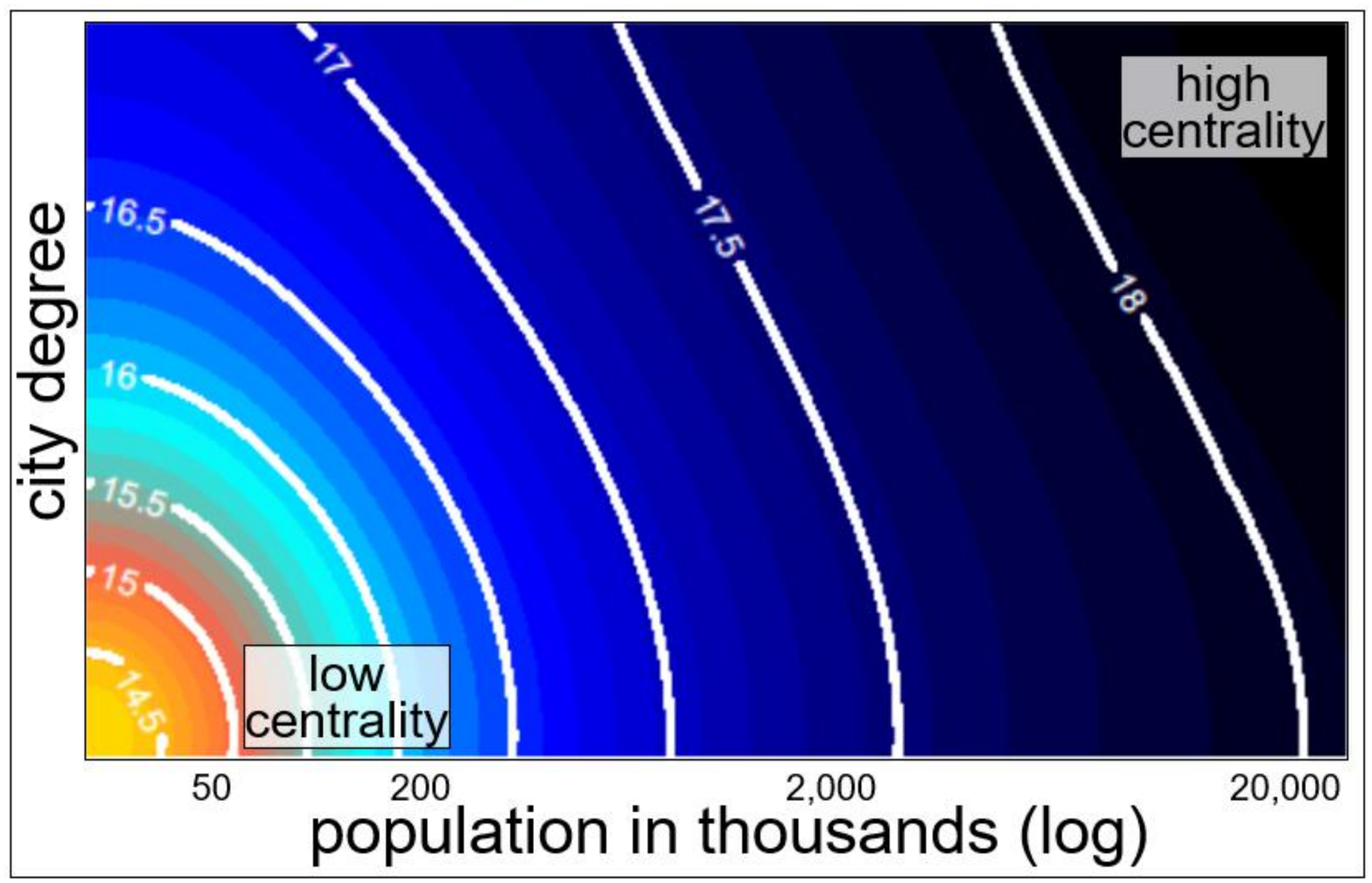}
\caption{Level of centrality according to city size (horizontal axis) and city degree (vertical axis).} 
\end{figure} \label{WhatIsHighCentral}
} 

Furthermore, cities in different regions tend to have different levels of centrality. Using cities from Central Africa as a baseline, we observe that in East Africa, for example, the centrality is higher (Table \ref{ModelCentralAfrica}).

\begin{table}[h!]
\centering
\begin{tabular}{r ll} 
Variable & coefficient & \\
\hline
Intercept & $0.62 \pm 0.18$ & $^{\ast \ast \ast}$ \\
\hline
Central Africa & - \\
East Africa & $0.34 \pm 0.23$ &\\
West Africa & $0.59 \pm 0.22$ &$^{\ast \ast}$\\
South Africa & $1.01 \pm 0.29$ &$^{\ast \ast \ast}$\\
North Africa & $1.83 \pm 0.22$ &$^{\ast \ast \ast}$ \\
\hline
population & $67.19 \pm 3.22$ & $^{\ast \ast \ast}$\\
population$^2$ & $31.86 \pm 3.12$ & $^{\ast \ast \ast}$\\
population$^3$ & $-7.61 \pm 3.1$ & $^{\ast}$\\
\hline
degree & $76.08 \pm 3.47$ &$^{\ast \ast \ast}$\\
degree$^2$ & $27.58 \pm 2.98$& $^{\ast \ast \ast}$\\
degree$^3$ & $-9.08 \pm 3.01$ &$^{\ast \ast}$\\
\hline
\end{tabular}
\caption{Coefficients of the regression with 2,162 observations (cities) using Central Africa as baseline. All values reported have $\times 10^6$ units. Adjusted R-squared: 0.4683}
\label{ModelCentralAfrica}
\end{table}

Our results show that North Africa has a dense and complex transport network, easing the flow of products and people between its cities and shortening distances by looking at the African continent's available road infrastructure. In particular, in Egypt, the high density of cities in the Nile delta means that not even high border costs significantly impact the country. However, the model also shows that in Central Africa, for example, the network is less dense and more simple, mainly connecting one city to another, with fewer crossings and bifurcations. Network distances are considerably larger, and some countries are strongly affected by the cost of borders. Thus, the centrality of its cities is significantly lower than in North Africa.

\section{Conclusions and discussion}

Africa is a vast, fragmented and sparsely populated territory. It is the least urbanised continent, and so the physical distance and the scarce road infrastructure impose high costs on trade and mobility and form a fragile network. Resources are limited, so prioritising investment to target and improve connections between cities is urgent. Our results shed light on how a continental-level transport network can be constructed, the flow between every possible origin and destination and the assumptions needed to capture the intensity of the flows. Measuring interactions between different cities is a challenge from many angles, including data scarcity, the high volume of variables and parameters, the wide range of regional, national and local conditions, among many. Some big data techniques could be applied to detect social contacts, travel, trade or other aspects to capture the impact of size and distance.

\subsection{The role of cities with high intermediacy}

There is an urgency to adapt national development plans as well as continental integration and corridor strategies to integrate better the central role of cities. There are thus two distinct challenges and opportunities for cities. Firstly, some secondary cities might be well-integrated within the urban network and might enjoy high centrality levels, suggesting that some industries might have a strategic potential in those locations, including all activities related to transportation, logistics and storage. On the opposite side of the spectrum, some secondary cities might be partly disconnected from the network or with low centrality levels. This aspect of cities is not due to the proximity to large cities, but rather, it comes from a much more complex relationship between cities which depends on the urban network. A scarce infrastructure and isolation due to the vastness of some countries and sparsely inhabited places, low-quality roads, high border costs or a poorly integrated system need to be addressed to spark the full potential of cities in a sustainable development and towards an integrated, prosperous and peaceful Africa.

\subsection{A model with assumptions and parameters}

Our method for constructing a connected road network depends on the data available, our assumptions and our procedure to connect cities and distinct parts of the network. More and better road data will improve the model results and improve the algorithm performance. Validating our results with aerial maps is almost an impossible task, taking into account the dimension of the problem. 

With little data to adjust our parameters, we modelled flow assuming values that work for all regions. With distinct data sources, it would be possible to test different values of the parameters and add non-linear effects of city size. Yet, it is possible that the impact of distance is even stronger in poorer parts of the continent and that trade and travels are less sensitive to distance in North Africa.

We ignored the competition costs between different users, but some of the primary roads are likely too congested, so travel times between cities are higher due to traffic. Likewise, we do not have information on road quality beyond the road type classification available in OSM. In some parts of Africa, poor road quality could imply that we significantly overestimate travel speeds. Also, we have ignored all maritime flow, which is not of negligible dimensions. The distance between two ports could be adjusted by considering marine travel times. Also, with naval or aerial routes, it would be possible to add Madagascar, Cape Verde and other islands into the model.

Although we considered all cities from continental Africa in the model, the impact of ignoring other parts of the world, particularly for our results in North Africa, is significant. We have ignored all trade and travels to Israel, Lebanon, Turkey, Greece, Cyprus, Italy, France, Spain and Portugal, but they are closer to Tunisia, for example than Namibia. Future works should aim at constructing a global network of cities, road infrastructure and flows at different geographical levels.

\newpage

\section{Supplementary information}

\subsection{Nodes in the urban network}

Describing the urban network in Africa requires two types of nodes: urban nodes or cities, which have a population larger than zero and a name identified by Africapolis \cite{Africapolis} and transport nodes. Transport nodes from the network correspond to road bifurcations, crossings or terminal nodes (so a node with degree one with no urban agglomeration nearby). Our simplification method has removed all transport nodes with degree two since they represent the ending of some road segment and the beginning of the next one, many of them corresponding to the border between two neighbouring counties or provinces (Figure \ref{TransportNodes}).

{
\begin{figure} \centering
\includegraphics[width=0.7\textwidth]{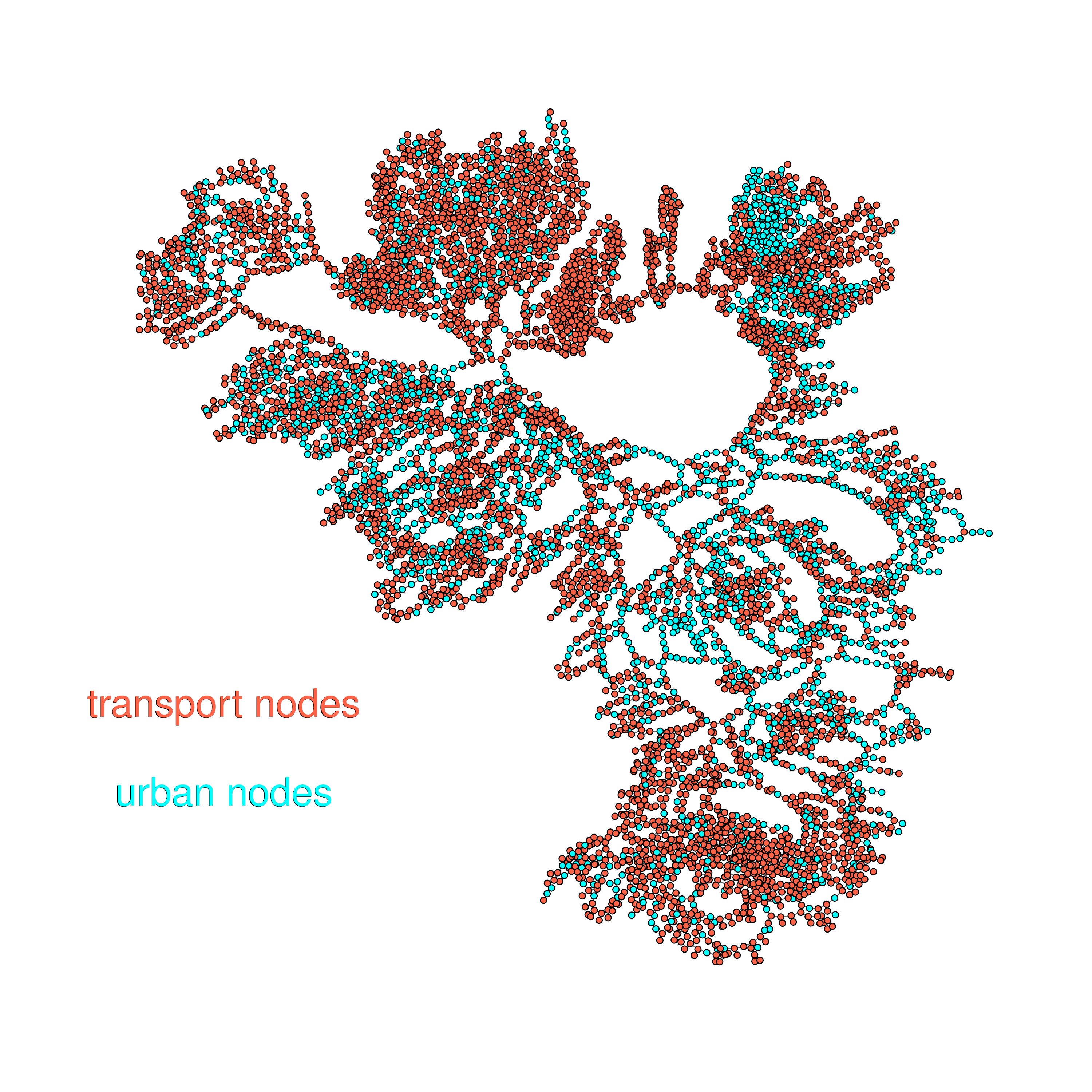}
\caption{Transport nodes are identified in red and cities in blue. In North Africa, many more transport nodes are observed, indicating a more complex road network.} \label{TransportNodes}
\end{figure}
} 

More dense and complex transport networks are observed in Egypt and South Africa, with many transport nodes and cities with a high node degree (Figure \ref{NodeDegreeMap}). 

{
\begin{figure} \centering
\includegraphics[width=0.7\textwidth]{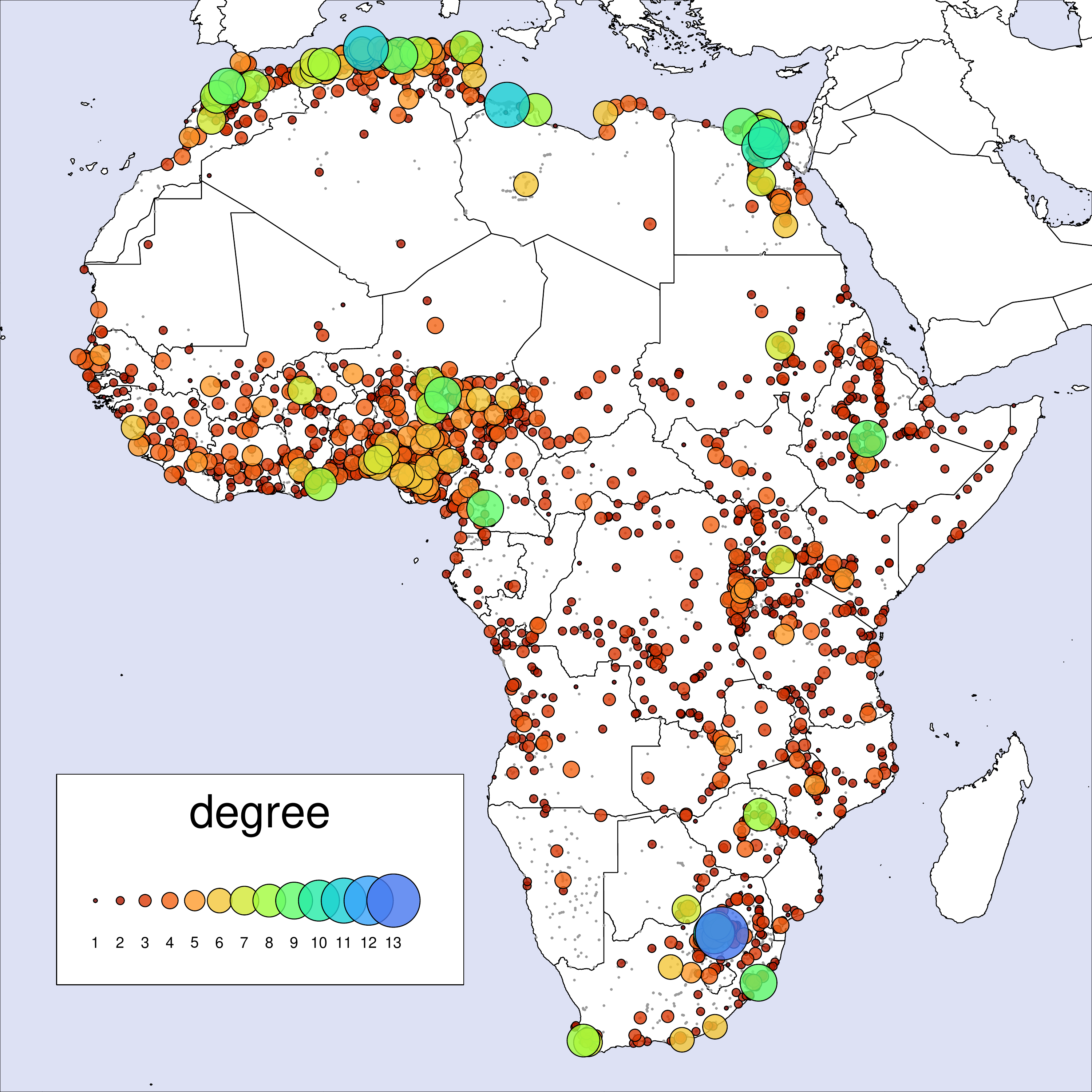}
\caption{The degree of each city corresponds to the colour. The size of the disc is proportional to city size. Small corridor cities are red nodes.} \label{NodeDegreeMap}
\end{figure}
} 

\subsection{Edges in the urban network}

The connected and simplified continental network of Africa obtain with our procedure can be described with 7,361 nodes (2,162 are cities and 5,199 are transport nodes, expressing road crossings and bifurcations) and with 9,159 edges.

We consider pairs of cities at distances smaller than 24km and compare the network and spatial distances. For cities with a ratio greater than 10 we added that edge. The reasoning behind this is that such small distances can be crossed by alternative methods even if OpenStreetMap data did not identify them. Likely, these roads correspond to secondary roads or other types of infrastructure beyond highways. Although there might be some errors here, where connections that do not exist are introduced, it reduces a more severe issue: seemingly disconnected patches of nearby nodes. Since we use the node betweenness as a centrality metric, the result is sensitive to a few disconnections, and so we keep this error to a minimum. In total, 1,675 edges were added (so 18\% of the edges) to obtain a fully connected network with reasonable travel times (Figure \ref{AddedEdgesMap}). 

{
\begin{figure} \centering
\includegraphics[width=0.7\textwidth]{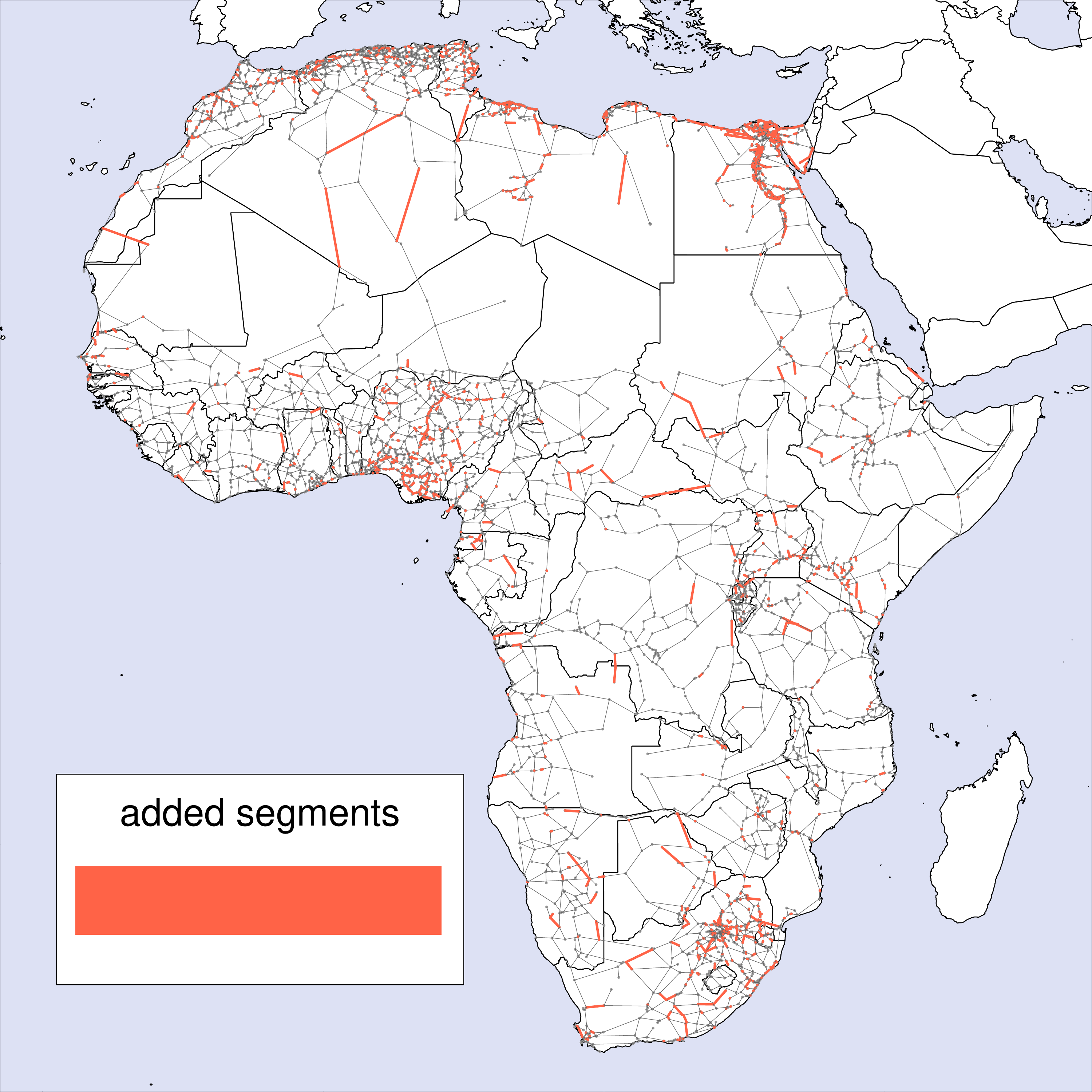}
\caption{Segments which were added to the network to get a connected and regular network are coloured in red. Other edges, obtained from OSM are gray.} \label{AddedEdgesMap} 
\end{figure}
}

We identify 299 border segments, where the extremes are in different countries, so only 3.3\% of the edges (Figure \ref{BorderEdgesMap}).

{
\begin{figure} \centering
\includegraphics[width=0.7\textwidth]{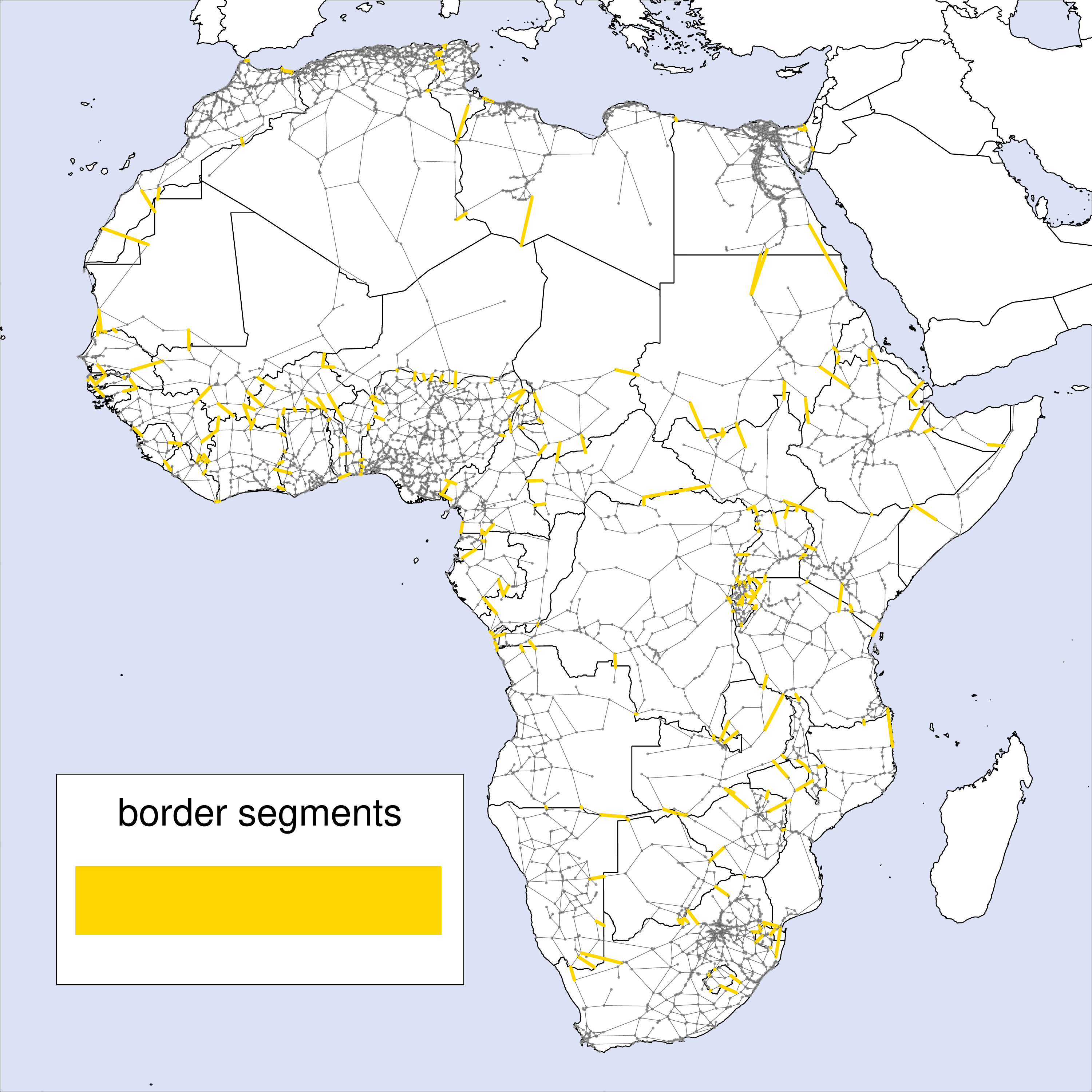}
\caption{Full network, where border segments are yellow. From the 9159 edges in the network, 299 cross a border (only 3.3\% of the edges in the continent).} \label{BorderEdgesMap}
\end{figure}
}

Edges with any of its extremes as a city are identified as urban. In total, 4266 segments are urban, 46.6\% as having one or two of their endings as urban. The rest of the edges (53.4\%) connect only transport nodes (Figure \ref{UrbanEdges}).

{
\begin{figure} \centering
\includegraphics[width=0.7\textwidth]{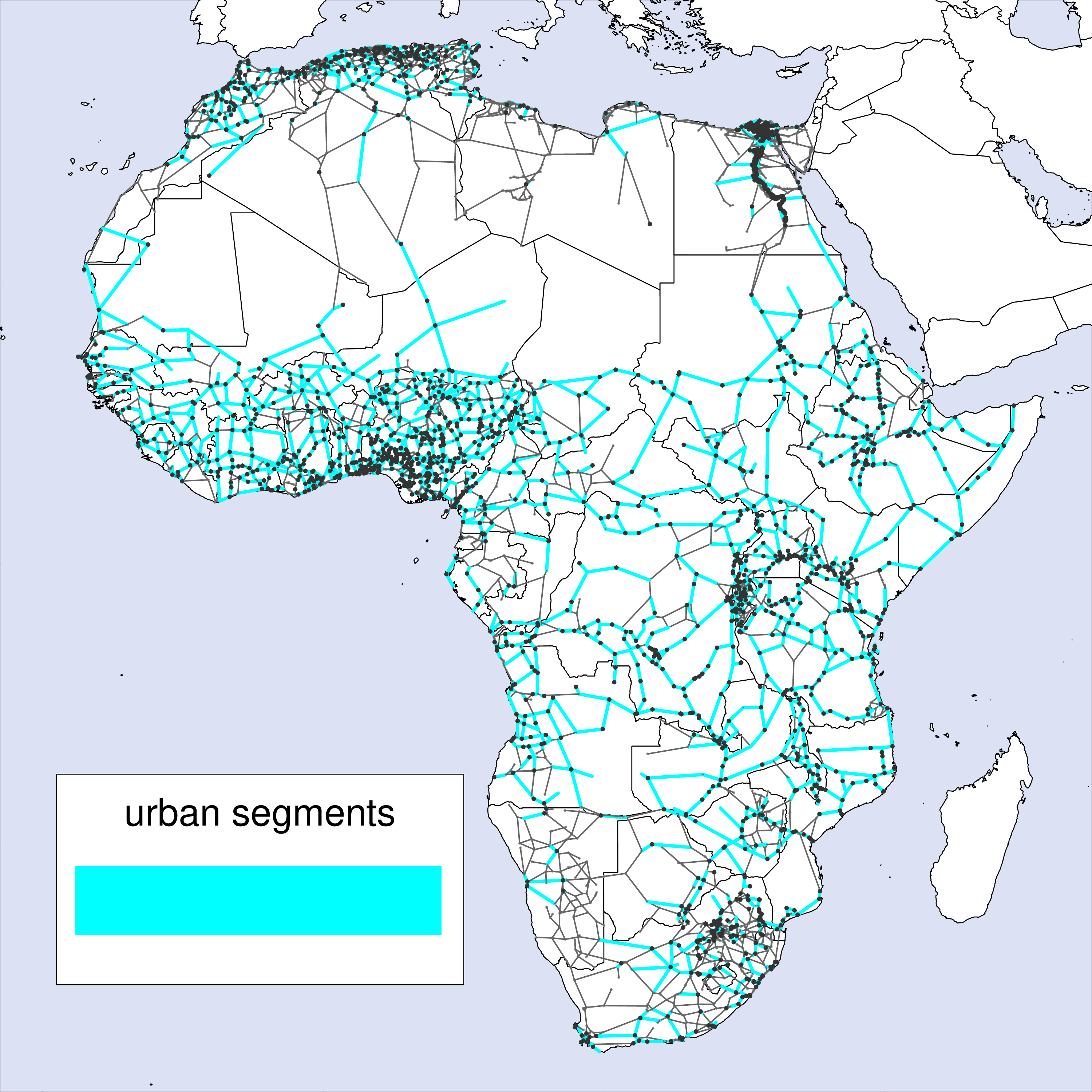}
\caption{Edges where either the start or the end are urban are blue. The rest are grey. } 
\label{UrbanEdges}
\end{figure}
} 

\subsection{Total infrastructure per country}

In total, 361,037km of road infrastructure were identified in continental Africa. This count does not include any streets or avenues in cities but rather is the sum of the lengths of all intracity roads from OpenStreetMap (Figure \ref{LengthPerCountry}). The countries where most road infrastructure was identified are Nigeria (with 31,248km), Algeria (with 29,039km) and South Africa (with 26,961km).

{
\begin{figure} \centering
\includegraphics[width=0.6\textwidth]{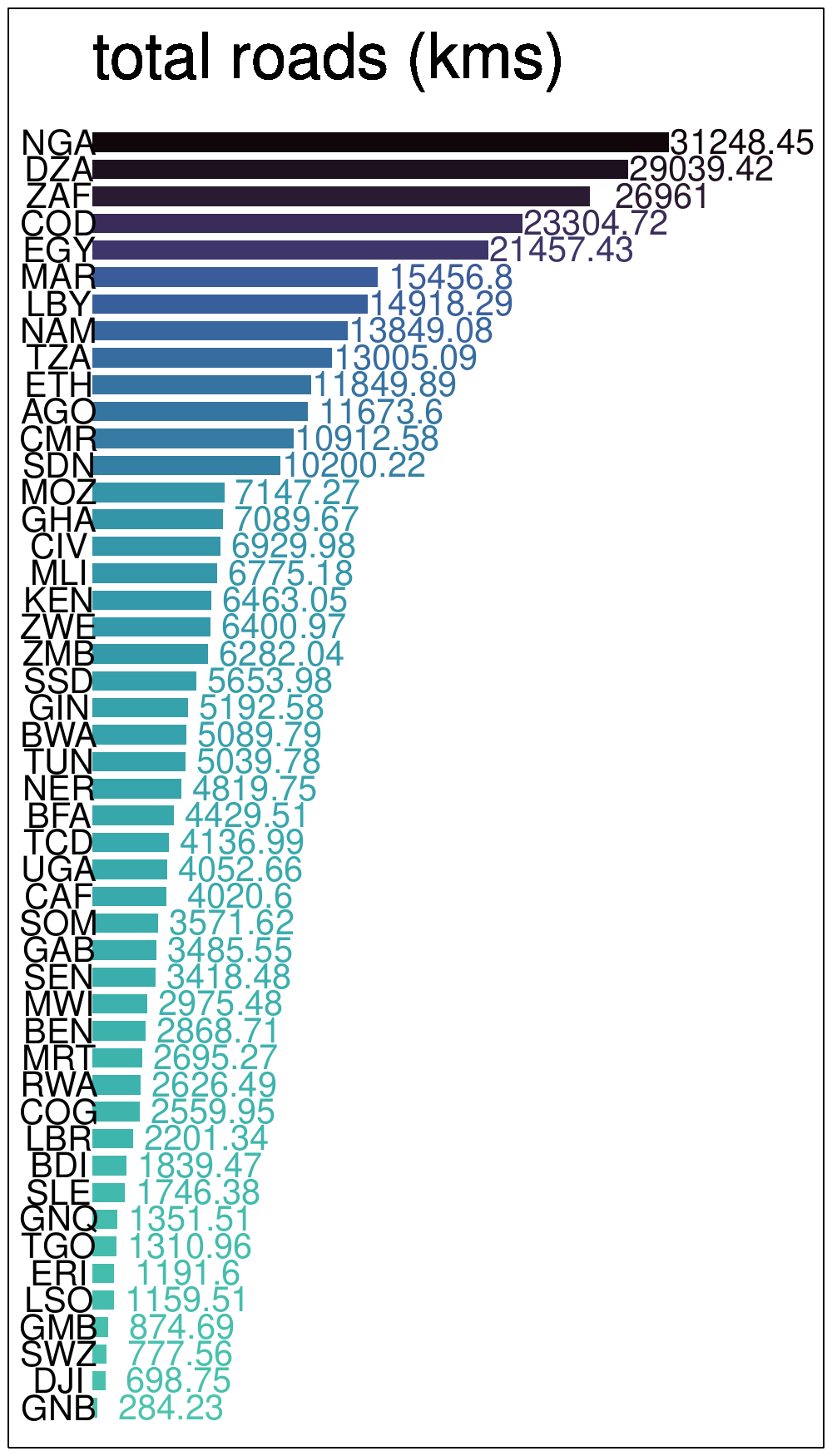}
\caption{Total length in kilometres for each country.} \label{LengthPerCountry}
\end{figure}
}

\subsection{Adjusting the cost of distance}

The gravity model has many expressions, and due to our limited city to city data, we have adopted the equation
\begin{equation}
F_{o,d} = \kappa \frac{P_o^\mu P_d^\nu}{D_{o,d}^\gamma},
\end{equation}
with the assumptions that $\kappa = 1$ and $\mu = \nu = 1$ so that the flow is symmetric and linear with respect to the size of origin and destination. Further, we use the network distance between two cities in the denominator and reduce the expression to a single parameter, $\gamma \geq 0$. We construct the cumulative percentage of trade that flows to a city at some distance threshold $d \geq 0$. We compare our modelled results to the ones observed in two West African cities (Ouagadougou and Tamale) \cite{FoodshedsTwoAfricanCities}. We use the average from the peak and the lean season and detect that values of $\gamma$ closer to 2 result in both cities having trade with cities that are too distant (Figure \ref{KargCumFigures}). For each city, we get the square difference between the observed and the modelled flows at ten different distance thresholds and minimise the sum of the errors for both cities. The value of $\hat{\gamma} = 2.8$ minimises the sum of the errors, showing that distance plays a significant burden in the observed flows.

{
\begin{figure} \centering
\includegraphics[width=0.7\textwidth]{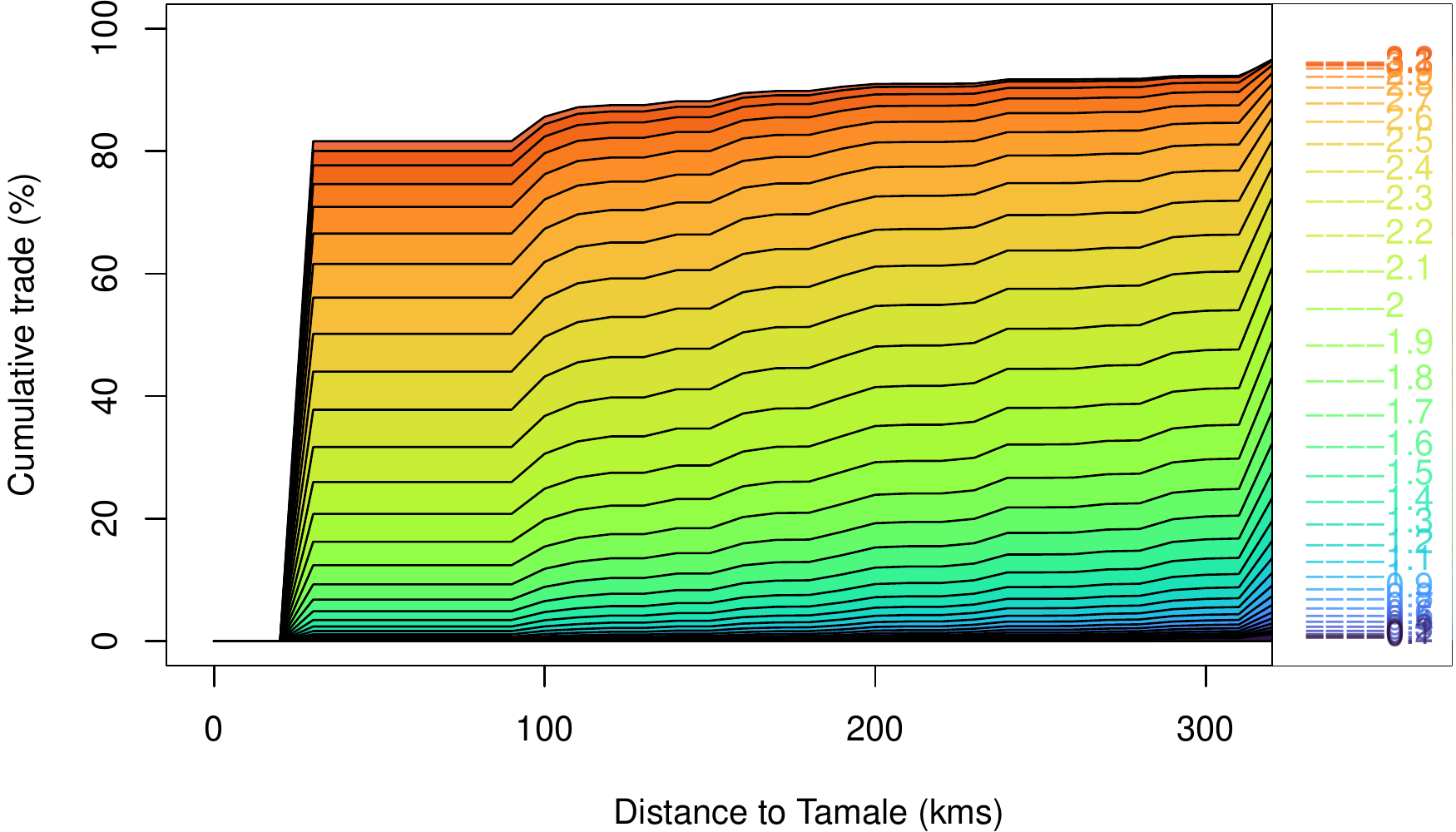}
\includegraphics[width=0.7\textwidth]{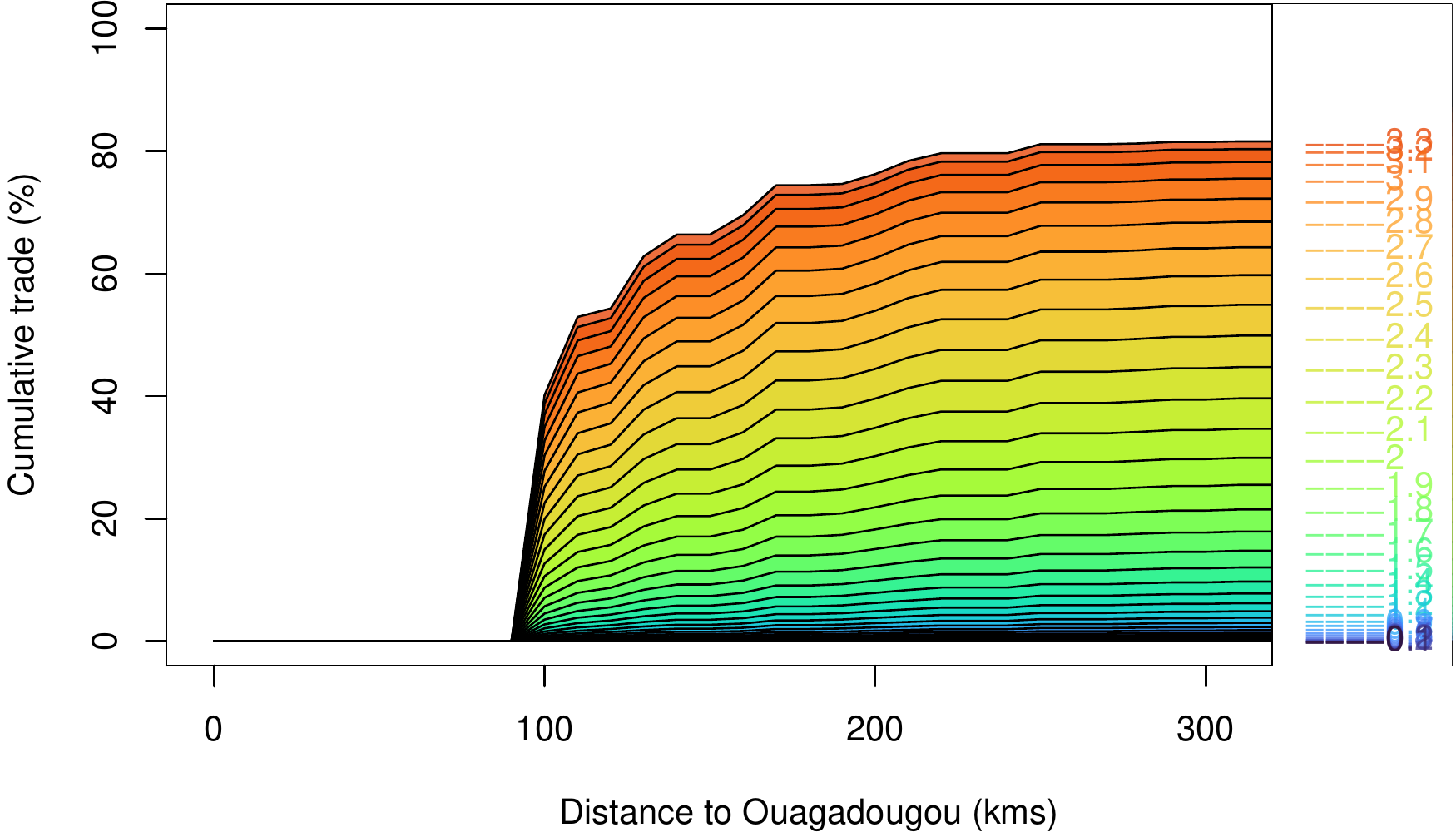}
\caption{Cumulative percentage trade (vertical axis) for Tamale and Ouagadougou that comes to each city from a distance $d$ (horizontal axis).} \label{KargCumFigures}
\end{figure}
}

\subsection{Border costs per region}

Depending on the interactions between countries and the relevance of borders, different regions suffer different impacts as the border cost $\tau$ increases. 

{
\begin{figure} \centering
\includegraphics[width=0.42\textwidth]{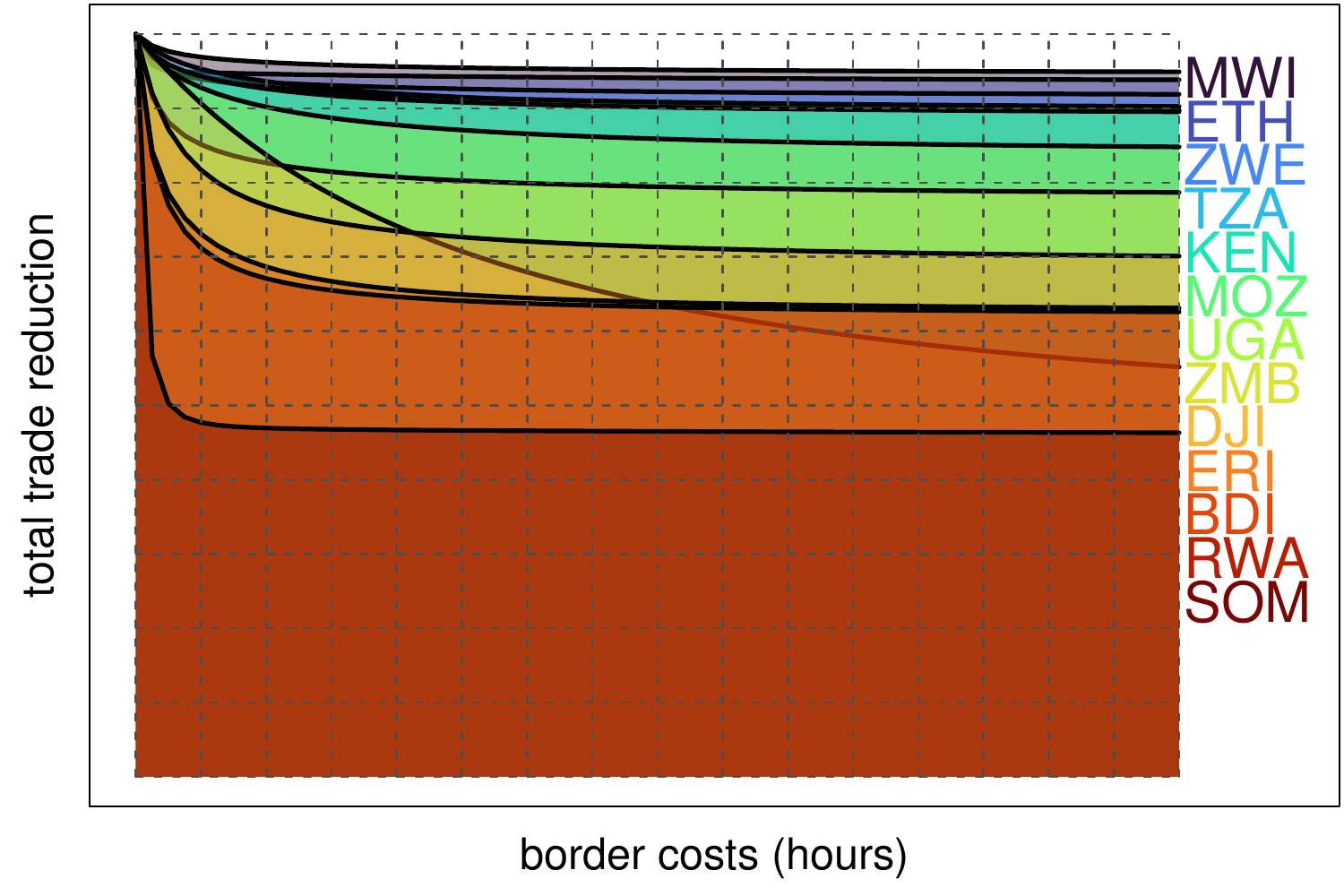}
\includegraphics[width=0.42\textwidth]{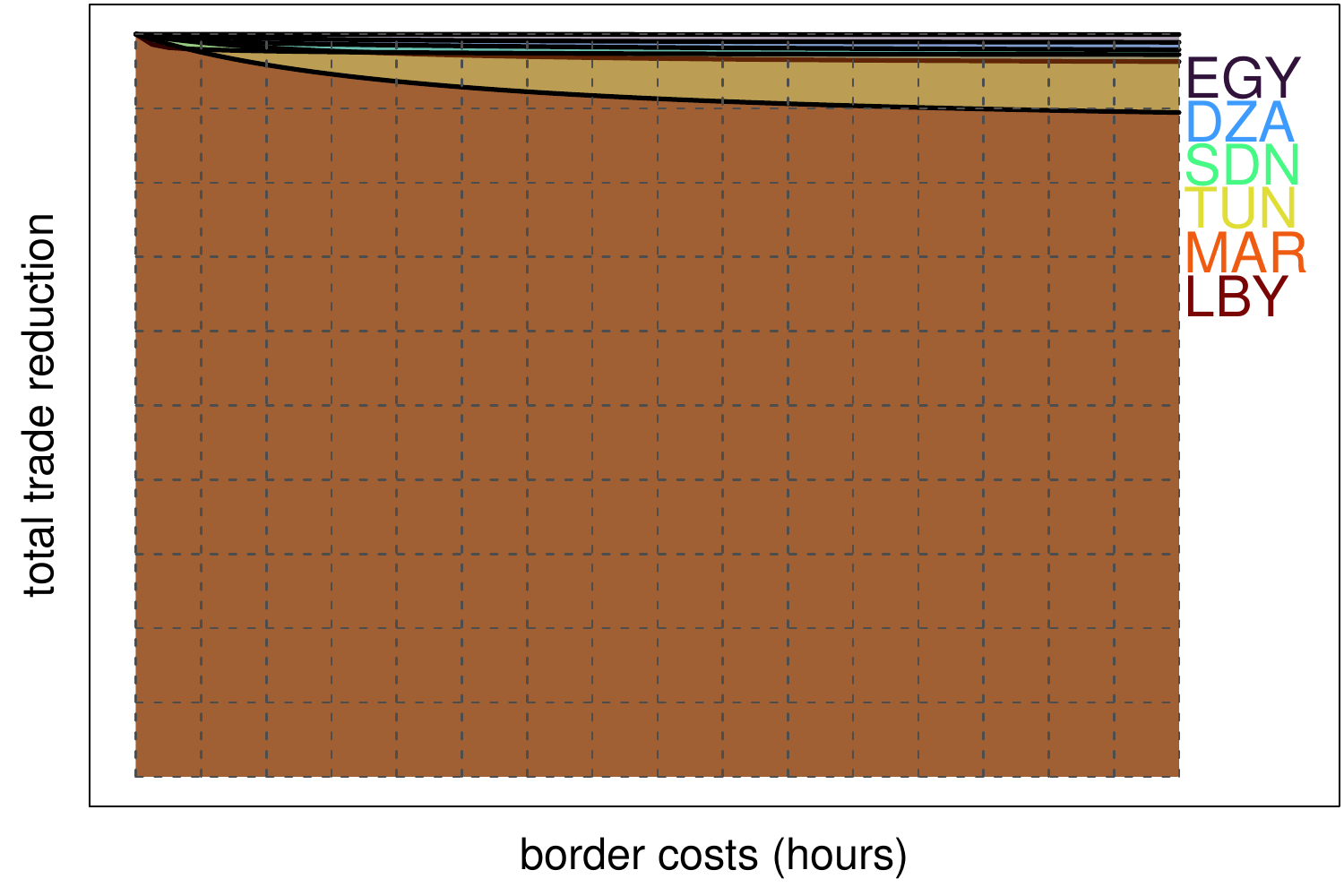}
\includegraphics[width=0.42\textwidth]{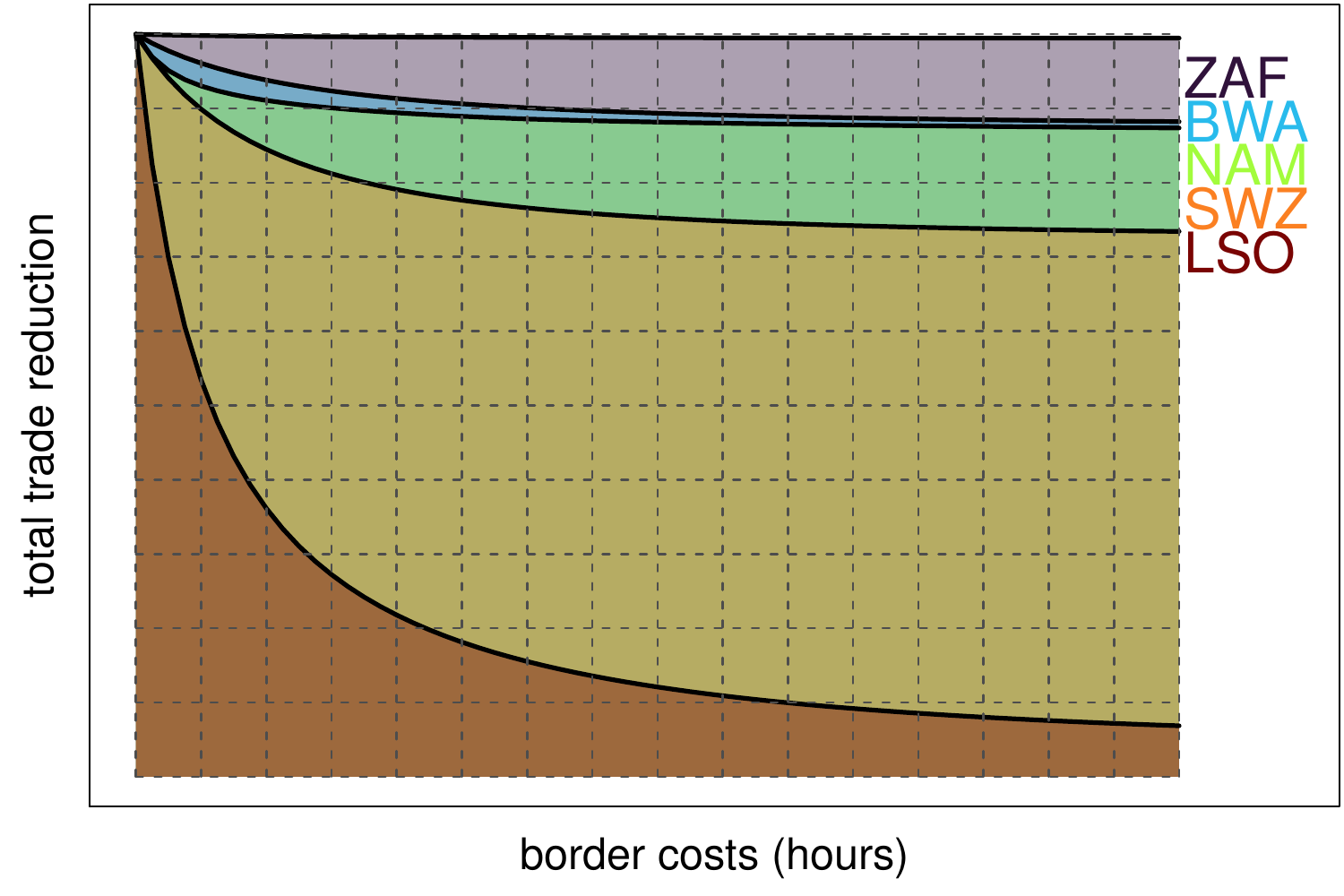}
\includegraphics[width=0.42\textwidth]{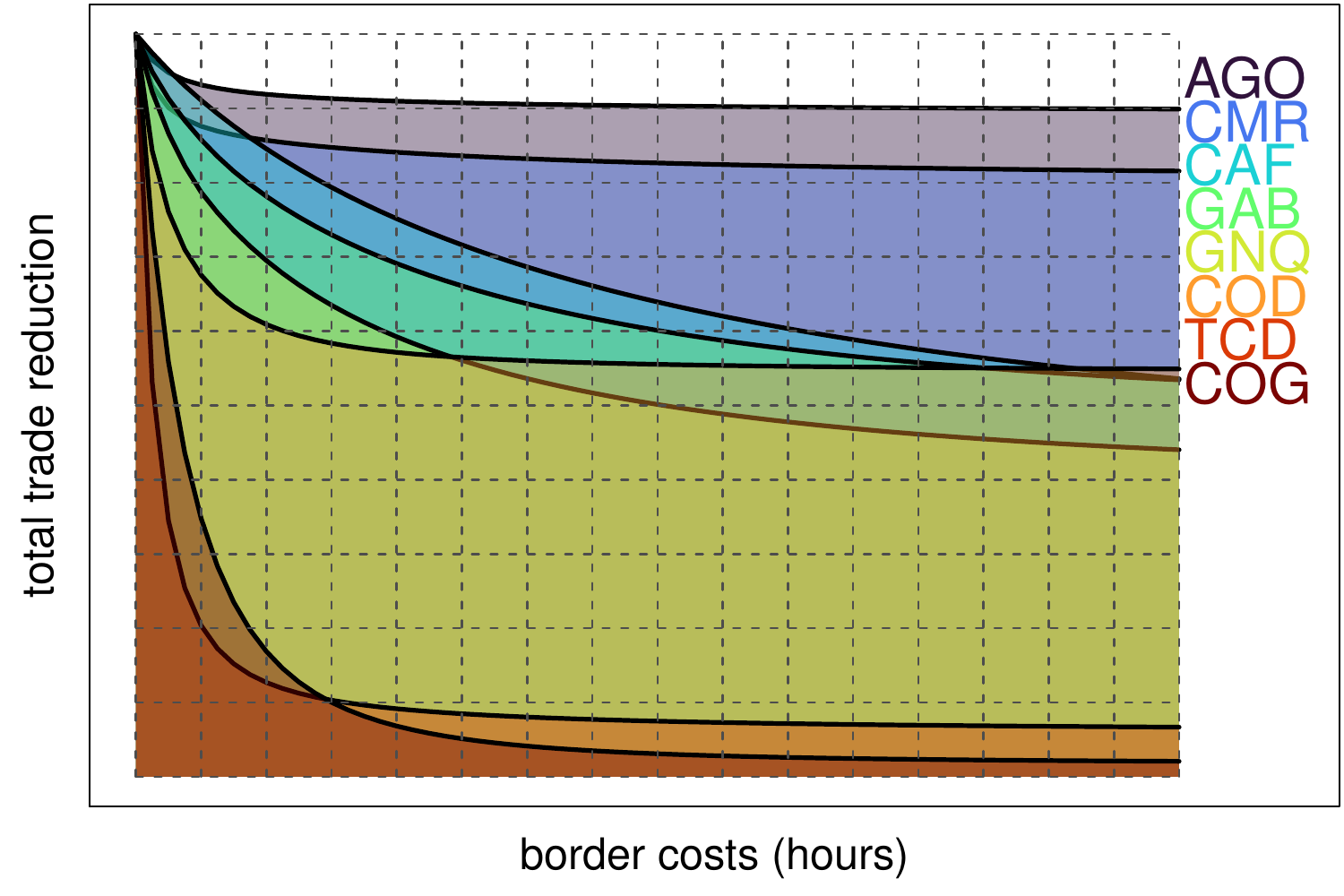}
\caption{Costs of international borders $\tau$, in hours (horizontal axis) against the ratio between the the total trade and the trade with no border costs (vertical axis) for West Africa.} \label{BorderCostsAll}
\end{figure}
}

\section{Materials and Correspondence}

All correspondence should be addressed to \textit{rafael.prieto.13@ucl.ac.uk}

\section*{Acknowledgments}

This article was completed with support from the PEAK Urban programme, funded by UKRI’s Global Challenge Research Fund, Grant Ref: ES/P011055/1.

Map data copyrighted OpenStreetMap contributors and available from https://www.openstreetmap.org.

\section*{Declaration of Competing Interest}

The authors declare that they have no known competing financial interests or personal relationships which have, or could be perceived to have, influenced the work reported in this article. 

\newpage

\bibliographystyle{unsrt}

\end{document}